\documentclass[11pt]{article}
\usepackage[sc]{mathpazo} 
\usepackage{fullpage}
\usepackage[authoryear,sectionbib,sort]{natbib}

\usepackage[utf8]{inputenc}
\usepackage{titlesec}
\titleformat{\section}[block]{\Large\bfseries\filcenter}{\thesection}{1em}{}
\titleformat{\subsection}[block]{\Large\itshape\filcenter}{\thesubsection}{1em}{}
\titleformat{\subsubsection}[block]{\large\itshape}{\thesubsubsection}{1em}{}
\titleformat{\paragraph}[runin]{\itshape}{\theparagraph}{1em}{}[. ]

\usepackage{amsmath}
\usepackage{graphicx}
\usepackage{dcolumn}
\usepackage{bm}
\usepackage[hidelinks]{hyperref}
\usepackage[dvipsnames]{xcolor}
\usepackage{palatino}
\usepackage{tcolorbox}
\usepackage{framed}
\usepackage{adjustbox}
\usepackage{tabularx}
\usepackage{booktabs}

\title{\vspace{-50pt}
Discovering stochastic dynamical equations from ecological time series data}

\author{Arshed Nabeel$^{1, 2, \ast}$ \\ 
Ashwin Karichannavar$^{2}$ \\ 
Shuaib Palathingal$^{2}$ \\
Jitesh Jhawar$^{3,4,5}$ \\
David B. Brückner$^{6, 7}$ \\
Danny Raj M$^{8, 9,\ast}$ \\
Vishwesha Guttal$^{2, \ast}$}

\newcommand{\bx}{\ensuremath{\mathbf x}}

\newcommand{\bz}{\ensuremath{\mathbf z}}
\newcommand{\bff}{\ensuremath{\mathbf f}}
\newcommand{\bg}{\ensuremath{\mathbf g}}
\newcommand{\bG}{\ensuremath{\mathbf G}}
\newcommand{\brr}{\ensuremath{\mathbf r}}
\renewcommand{\bm}{\mathbf{m}}
\newcommand{\modm}{|\bm|}
\newcommand{\boldeta}{\boldsymbol{\eta}}
\newcommand{\boldeps}{\boldsymbol{\varepsilon}}

\setcitestyle{authoryear,round}
\renewcommand{\cite}{\citep}

\date{}

\begin{document}

\maketitle

\noindent{} 1. IISc Mathematics Initiative, Indian Institute of Science, Bengaluru Karnataka, 560012, India

\noindent{} 2. Centre for Ecological Sciences, Indian Institute of Science, Bengaluru Karnataka, 560012, India

\noindent{} 3. University of Konstanz, Konstanz, Germany

\noindent{} 4. Max Planck Institute of Animal Behaviour, Konstanz, Germany.

\noindent{} 5. School of Arts and Sciences, Ahmedabad University, Ahmedabad, 380009, India. 

\noindent{} 6. Institute of Science and Technology, Austria, Am Campus 1, 3400 Klosterneuburg, Austria

\noindent{} 7. Biozentrum, University of Basel, Switzerland

\noindent{} 8. Department of Chemical Engineering, Indian Institute of Science, Bengaluru, Karnataka, 560012, India 

\noindent{} 9. Dept of Applied Mechanics and Biomedical Engineering, IIT Madras, Chennai, 600036, India

\noindent{} $\ast$ Corresponding authors' e-mail: arshed@iisc.ac.in, danny@iitm.ac.in and guttal@iisc.ac.in

\bigskip

\textit{Keywords}: Data Driven Model Discovery, Langevin Dynamics, Self-organization, Collective motion, Mesoscale dynamics, Data Driven Dynamical Systems, Scientific Machine Learning, Noise induced order.


\newpage{}

\section*{Abstract}
Theoretical studies have shown that stochasticity can affect the dynamics of ecosystems in counter-intuitive ways. However, without knowing the equations governing the dynamics of populations or ecosystems, it is difficult to ascertain the role of stochasticity in real datasets. Therefore, the inverse problem of inferring the governing stochastic equations from datasets is important. Here, we present an equation discovery methodology that takes time series data of state variables as input and outputs a stochastic differential equation. We achieve this by combining traditional approaches from stochastic calculus with the equation-discovery techniques. We demonstrate the generality of the method via several applications. First, we deliberately choose various stochastic models with fundamentally different governing equations; yet they produce nearly identical steady-state distributions. We show that we can recover the correct underlying equations, and thus infer the structure of their stability, accurately from the analysis of time series data alone. We demonstrate our method on two real-world datasets – fish schooling and single-cell migration – which have vastly different spatiotemporal scales and dynamics. We illustrate various limitations and potential pitfalls of the method and how to overcome them via diagnostic measures. Finally, we provide our open-source codes via a package named \textbf{PyDaDDy} ({\bf Py}thon library for \textbf{Da}ta \textbf{D}riven \textbf{Dy}namics).
\newpage{}

\section*{Introduction}
A central approach to modeling complex ecological systems across fields is via differential equations~\cite{strogatz2018nonlinear,gotelli2008primer}. Depending on the dimensionality of the system and the relevance of stochasticity, these equations can be ordinary, stochastic, or partial differential equations. Even when we begin with a simple set of behavioral rules or ecological interactions at a fine, local scale, one can derive coarse-grained dynamical descriptions, at levels such as groups, populations or even ecosystems, in the form of differential equations ~\cite{mckane2004stochastic,cheng2014mesoscopic, yates2009locust, 
loreau2010populations,durrett1994tpb,biancalani2014prl,jhawar2019bookchapter,majumder2021finite}. 
While such dynamical system based approaches are powerful and have provided key biological insights, meaningfully integrating empirical data with differential equation models continues to remain a challenge. 

We are in the era of big data, where high-resolution data capturing the dynamics of ecological systems is increasingly available~\cite{leyk2019spatial,nathan2022big}. Examples span across scales of biological organization, from individual (cells or animals) movement trajectories~\cite{bruckner2019stochastic,nathan2022big} and group properties~\cite{yates2009locust,jhawar2020fish,tunstrom2013collective} to population sizes~\cite{stenseth1997population,bjornstad2001noisy}, fitness of populations~\cite{lenski2017experimental} and ecosystem states~\cite{carpenter2020stochastic,xie2008remote,majumder2019inferring}. Importantly, data may be available with high temporal resolutions, sometimes even across space, thus opening avenues for better integration between models and data. 

To capture the dynamics of biological systems accurately, one has to treat state variables as nonlinear and stochastic, rather than accounting only for the linear and average properties.  A suitable framework for analyzing such myriad stochastic effects is via \emph{stochastic differential equations (SDEs)}, which allows us to capture both the deterministic and stochastic factors driving the dynamics. Using SDE models, the most commonly understood effect of noise is that of creating fluctuations around a deterministic stable state. This resembles a population fluctuating around its carrying capacity equilibrium due to environmental noise. However, the SDE models also predict that when the strength of the noise depends on the state of the system -- also called state-dependent noise -- it can create new states away from the deterministic stable equilibria~\cite{horsthemke1989book}. For example, in fish schools, fluctuations associated with small group sizes, counter-intuitively, push the system away from deterministically stable disordered states; consequently, the group level coordination among fish increases~\cite{biancalani2014prl,jhawar2019bookchapter, jhawar2020fish}. In the context of dryland ecosystems, mathematical models predict that rainfall fluctuations may create a new state between two deterministic stable states~\cite{odorico2005pnas}; this is in contrast to well-known scenarios where additive fluctuations typically lead to transitions between alternative stable states~\cite{guttal2007ecomod}. However, these conclusions on the role of stochasticity are possible only when the governing dynamical equations are known, which is often not the case for real datasets. 

Hence, we naturally focus on the inverse problem: can we construct SDE models starting from observed time series datasets? Indeed, the answer is in the affirmative. Approaches based on estimating the so-called \emph{jump moments}, in principle, allow us to infer stochastic differential equations from time series data \cite{gradivsek2000analysis,friedrich2011approaching,tabar2019book, rinn2016langevin}. Furthermore, in the context of deterministic models, recently developed \emph{equation discovery} techniques allow us to infer parsimonious, interpretable differential equation models from time series data~\cite{brunton2016sindy, rudy2017pdefind, desilva2020pysindy}. Recently, these techniques have also been extended for stochastic dynamical systems~\cite{boninsegna2018sparse,huang2022sparse,callaham2021langevin,frishman2020sfi,bruckner2020inferring}. These approaches are promising for analysing ecological time series datasets. 

However, there are significant gaps. These techniques are scattered across physics and engineering literature; they are, thus, not well known in biological contexts. Importantly, there is no fundamental reason to believe that SDEs which assume relatively simple noise structures are reasonable models for real biological datasets. Therefore, proper diagnostic tools that test the assumptions of noise and validate the correct model are essential; however, such diagnostics are often not the focus in the physics and engineering literature where many new techniques are being developed. Consequently, the methods of stochastic equation discovery are currently not readily amenable for biological applications.

In this manuscript, we bridge these gaps and present a unified framework for discovering and diagnosing SDE models for biological datasets. Put simply, our approach allows a user to input a time series and discover an underlying stochastic differential equation and then perform diagnostics to test the validity of the discovered model. We argue that this approach is novel and powerful for several reasons: First, we arrive at a parsimonious and interpretable stochastic dynamical system (SDE) model directly from the data. Second, deciphering the function involves analysis of stochastic fluctuations of the input time series at the finest available time scales; yet the discovered dynamical model can correctly capture the long-time scale features of the data. We emphasize that these long-term features were not used as input to the model discovery. Third, the equation learning \emph{discovers} an appropriate functional form of the model of the given dynamical dataset with minimal input from the user; and not merely estimate the parameters of a user-defined function. 

To illustrate the generality of the method,  we consider several applications: (A) We deliberately choose SDEs where deterministic and stochastic terms are fundamentally different; however, they produce very similar steady-state distributions. (B) Classic ecological models with stochasticity. (C) Two previously published real-world datasets having contrasting scales and dynamics: (i) a schooling fish movement driven by stochasticity and (ii) a single cell movement driven by a deterministic limit cycle with a minimal role of stochasticity. We extensively discuss various limitations and how the diagnostic tools can be used to avoid potential pitfalls and to test whether the discovered models meet underlying assumptions and accurately describe the data. Finally, to make our approach widely accessible, we provide our open-source codes for the analysis of one/two dimensional datasets via a package named \textbf{PyDaDDy} ({\bf Py}thon library for \textbf{Da}ta \textbf{D}riven \textbf{Dy}namics)~\cite{pydaddy-zenodo}, available at \url{https://github.com/tee-lab/PyDaddy} (archived at \url{https://doi.org/10.5281/zenodo.13777396}~\cite{pydaddy-zenodo}).  

\section*{Methods}
\subsection*{Mathematical preliminaries of stochastic differential equations}
Our goal is to model the dynamics of systems starting from measurements of the system state observed as a time series. Examples include the population sizes of one or multiple interacting populations, trajectories of foraging animals, vegetation cover in a landscape, etc. To do so, we use the framework of stochastic differential equations (SDE). Let $\bx(t)$ be a $d$-dimensional vector. We model its temporal dynamics as 
\begin{align}
    \dot \bx = \bff(\bx) + \bg(\bx) \boldeta(t) \label{eq:sde}
\end{align}
which is interpreted in an {\it It\^{o} sense}~\cite{van1992stochastic}. Here, $\bff$ and $\bg$ are functions of $\bx$, and $\boldeta(t)$ is a $d$-dimensional Gaussian white-noise~\cite{gardiner2009}, uncorrelated across components and across time (i.e, $\langle \boldeta_i \boldeta_j \rangle = \delta_{ij}$ and $\langle \boldeta(t) \boldeta(t')^T \rangle = \delta(t-t') I_{d \times d}$.

Here, $\bff(\bx)$ is a $d$-dimensional vector function, called the deterministic or \emph{drift} function, governing the average rate of change of $\bx$ over time. $\bg$ is a $d \times d$ matrix function, called the stochastic or \emph{diffusion} function, and governs the fluctuations around the average value. The off-diagonal entries of $\bg$ capture correlated fluctuations across components.

When $\bg$ is a constant matrix independent of $\bx$, the strength of the noise is the same for all values of $\bx$, and is called \emph{additive} or \emph{state-independent} noise. When $\bg$ is a function of $\bx$, the strength of the noise depends on the instantaneous value of $\bx$, and is called \emph{multiplicative} or \emph{state-dependent} noise~\cite{horsthemke1989book}.

If $\bx(t)$ is governed by the SDE in Eq.~\ref{eq:sde}, the drift and diffusion functions can be expressed in terms of the \emph{Kramers-Moyal coefficients} as follows:

\begin{align}
    \bff(\bz) &= \lim_{\Delta t \to 0} \left \langle \frac{\bx(t + \Delta t) - \bx(t)}{\Delta t} \right \rangle_{\bx(t) = \bz}\\
    \mathbf{G}(\bz) &= \lim_{\Delta t \to 0} \left \langle \frac{(\bx(t + \Delta t) - \bx(t))(\bx(t + \Delta t) - \bx(t))^T}{\Delta t} \right \rangle_{\bx(t) = \bz}
\end{align}

where $\mathbf{G} = \bg \bg^T$.

The following are the assumptions under which a given stochastic process $\bx$ can be modelled using a stochastic differential equation. First, $\bx(t)$ should be a \emph{continuous} stochastic process or should be approximated by one. Second, the noise $\boldeta(t)$ is assumed to be uncorrelated across time. Note that the assumption of the components of $\eta$ being uncorrelated is without loss of generality: if $\boldeta(t)$ had a non-diagonal correlation matrix $L$, an SDE $\bx(t) = \bff(\bx) + \bg(\bx) \boldeta(t)$ driven by $\boldeta(t)$ can be replaced by an equivalent SDE, $\bx(t) = \bff(\bx) + \tilde \bg(\bx) \boldsymbol{\zeta}(t)$ where $\boldsymbol{\zeta}$ is uncorrelated and $\tilde \bg(\bx) = \bg(\bx) \sqrt{L}$.  Here, $\sqrt{L}$ is the (not necessarily unique) matrix square root such that $\sqrt{L}\sqrt{L}^T = L$.

\subsection*{The basic principle behind the inverse problem}

We are interested in the inverse problem of inferring SDE from the observed time series data, sampled with some finite sampling frequency. Specifically, we aim to find simple, interpretable analytical expressions that describe $\bff$ and $\bG$, not just their qualitative shapes.

Our approach to this problem involves two steps. First, we compute instantaneous estimates of the drift and diffusion functions from the observed time series. Next, we use a technique based on sparse regression to estimate functional forms for the drift and diffusion functions.

\paragraph{Instantaneous drift and diffusion coefficients} 
From an observed sampled time series with some finite sampling time $\Delta t$, we can estimate the instantaneous drift coefficient as:
\begin{align}
    F(t; \bx) &= \frac{\bx(t + \Delta t) - \bx(t)}{\Delta t} \label{eq:instadrift}
\end{align}
Here, $F(t; x)$ is an instantaneous estimate for the drift coefficient, computed at every time step. Which will be subsequently used to estimate an expression for the drift function (see below). 

One can estimate the instantaneous diffusion coefficient $G(t; \bx)$ similarly. To improve the estimate of $G(t; \bx)$ and to counter the effect of a finite $\Delta t$, we compute the \emph{residuals} after subtracting the expected deterministic drift in time $\Delta t$ and estimate the diffusion from these residuals~\cite{jhawar2020inferring, bruckner2019stochastic}.

\begin{align}
    \boldeps(t) &= \bx(t + \Delta t) - \bx(t) - \Delta t \, \hat \bff(\bx(t)) \\
    G(t; \bx) &= \frac{(\boldeps(t + \Delta t) - \boldeps(t))(\boldeps(t + \Delta t) - \boldeps(t))^T}{\Delta t}  \label{eq:instadiff}
\end{align}

Here, $\hat \bff(\bx(t))$ is the estimated drift function (see below).

\paragraph{Equation learning for the drift and diffusion coefficients} We use a technique based on sparse regression, sometimes referred to as \emph{equation learning}, to find interpretable analytical expressions of the extracted drift and diffusion functions~\cite{brunton2016sindy, boninsegna2018sparse}. Without further assumptions, this can be done component-wise for each component of the $\bff$ and $\bG$ separately. To do so, we first construct a library $\{f_1, f_2, \ldots f_k\}$ of basis functions. These functions could be a polynomial basis up to a specified degree, another suitable basis such as a Fourier or Chebyshev basis, or domain-specific basis functions tailored for the problem. Next, we use sparse regression to represent $f$ as a linear combination of these basis functions, using as few $f_i$'s as possible. Operationally, this is done using an algorithm called \emph{sequentially thresholded least squares} (STLSQ; see SI Section S1 A - \emph{Sparse regression with sequentially thresholded least-squares} for a summary, based on~\cite{brunton2016sindy}). 

\subsection*{Model selection} 
Apart from the basis function, the equation discovery algorithm requires additional input from the user that constraints the model selection: the sparsification threshold $\lambda$. The value of $\lambda$ determines the trade-off between model complexity and explanatory power: a higher value of $\lambda$ will eliminate more terms to produce a simpler model, at the cost of a poorer fit to data. The value of $\lambda$ can be manually set to optimize this trade-off for the problem, or can be automatically tuned for. One could use an information criterion like the Akaike Information Criterion (AIC) or Bayesian Information Criterion (BIC) for the threshold selection. Here, we instead use \emph{$k$-fold cross validation}, a method commonly used in machine learning literature to choose the ideal model based on cross-validated accuracy~\cite{shalev2014understanding}. 

The key idea is to train a model using only part of the data, called \emph{training set}, and evaluate the model performance on a \emph{validation set}, different from the training set. A model that performs well on the validation set will not have overfit on the noise in the training data, and may be more generalisable. Specifically, the dataset is divided into $k$ equal chunks. $k - 1$ of the chunks are used as the \emph{training set} to fit the model, with the remaining chunk designated as the \emph{validation set}. The error of the model on the validation set, called the \emph{validation error}, is then computed. The process is repeated with each of the $k$ chunks designated as the validation set, and the average validation error is computed. We choose the optimal threshold the value that offers the \emph{maximum drop} in CV error. For many systems, including real-world systems, it may be possible to incorporate knowledge about the physics/biology of the system to choose the basis functions as well as do a manual model selection, which can result in a better and more parsimonious model (see~\citep{nabeel2023data} and SI Section S4 -- \emph{Model selection with real-world datasets}). 

\subsection*{Diagnostics}
The data-driven SDE discovery pipeline is incomplete without sufficient diagnostic tests to reasonably satisfy the underlying assumptions of the SDE models. Extending ideas from~\cite{bruckner2019stochastic,jhawar2020inferring}, we propose three sets of diagnostic tests that enable evaluation of the discovered SDE model.

\begin{itemize}
    \item \emph{Noise diagnostics:} The noise term $\boldeta(t)$ in Eq.~\ref{eq:sde} is assumed to be uncorrelated Gaussian noise. 
    We define the model residual as, 
    \begin{align}
        \brr(t) = \frac{1}{\sqrt{\hat{\bG}(\bx(t))}} \left( \frac{\bx(t + \Delta t) - \bx(t)}{\Delta(t)} - \hat{\bff}(\bx(t)) \right)
    \end{align}
    where $\hat \bff$ and $\hat \bG$ are the estimated drift and diffusion functions, and the square root, in general, is a matrix square root. $\brr(t)$ is an estimate of $\boldeta(t)$. We therefore test if $\brr(t)$ is an uncorrelated Gaussian noise. More advanced diagnostic tests to check noise assumptions are described in detail in SI Section 1B (\emph{Advanced diagnostic techniques}).

    \item \emph{Model diagnostics:} The idea of model diagnostics is that if we generate a simulated time series using the estimated model, with identical length and sampling interval as the original time series, then the simulated time series should have the same probability distribution and autocorrelation function as the original time series. We note that these two properties of the data were not used as part of the data-driven model discovery procedure. Further, if we subsequently estimate an SDE model from this simulated time series, we must recover the same model (also known as `model self-consistency'). 

    \item \emph{Validation with left-out data:} To check if the discovered model is sufficiently general, we divide the datasets into two halves, a \emph{training set} and a \emph{validation set} (also called {\it left-out data}) and perform the model discovery procedure using only the training set. We can now compute the histogram and autocorrelation for a time series generated by the discovered model and compare these with the histogram and autocorrelation of the left-out data. If the model statistics match well with the left-out data, it suggests that the discovered SDE model is generalizable.
    
\end{itemize}


\subsection*{Contrast to conventional approaches}


A conventional approach often used in literature to recover SDEs is to divide the range of $\bx$ into a finite number of bins and estimate $\bff$ and $\bG$ as bin-wise averages. That is, the drift and diffusion functions can be approximated from their instantaneous counterparts (Eqs.~\ref{eq:instadrift}, \ref{eq:instadiff}) as:

\begin{align}
    \hat \bff(\bz) = \langle F(t; \bx) \rangle_{\bx(t) \in [\bz, \bz + w]} \\
    \hat \bG(\bz) = \langle G(t; \bx) \rangle_{\bx(t) \in [\bz, \bz + w]} \\
\end{align}

Here, $w$ is the bin-width, and the intervals are $d$-dimensional rectangular intervals. The drawback of this approach is that the estimate for each bin is independent and cannot take advantage of any global structure, such as the smoothness of the functions involved. As a result, the estimates can be noisy. One way to reduce the variance of the estimates is to subsample the time series with a larger $\Delta t$~\cite{jhawar2020inferring}, but this can introduce biases in the estimation (see SI Section S5 A -- \emph{Estimation with limited data}).

In contrast, the equation learning approach replaces the bin-wise averaging step with a sparse regression step. In addition to the obvious advantage of producing interpretable expressions as estimates, equation learning has two additional advantages. First, by appropriately choosing the basic functions for the library, we can account for the global structure or constraints such as smoothness and symmetry. In addition, the need to subsample is eliminated: thus, two arbitrary parameter choices, namely for the subsampling time $\Delta t$ and the bin width $\epsilon$ are eliminated.

\subsection*{PyDaDDy -- an open-source package for data-driven SDE discovery}

\begin{figure*}
    \includegraphics[width=\textwidth]{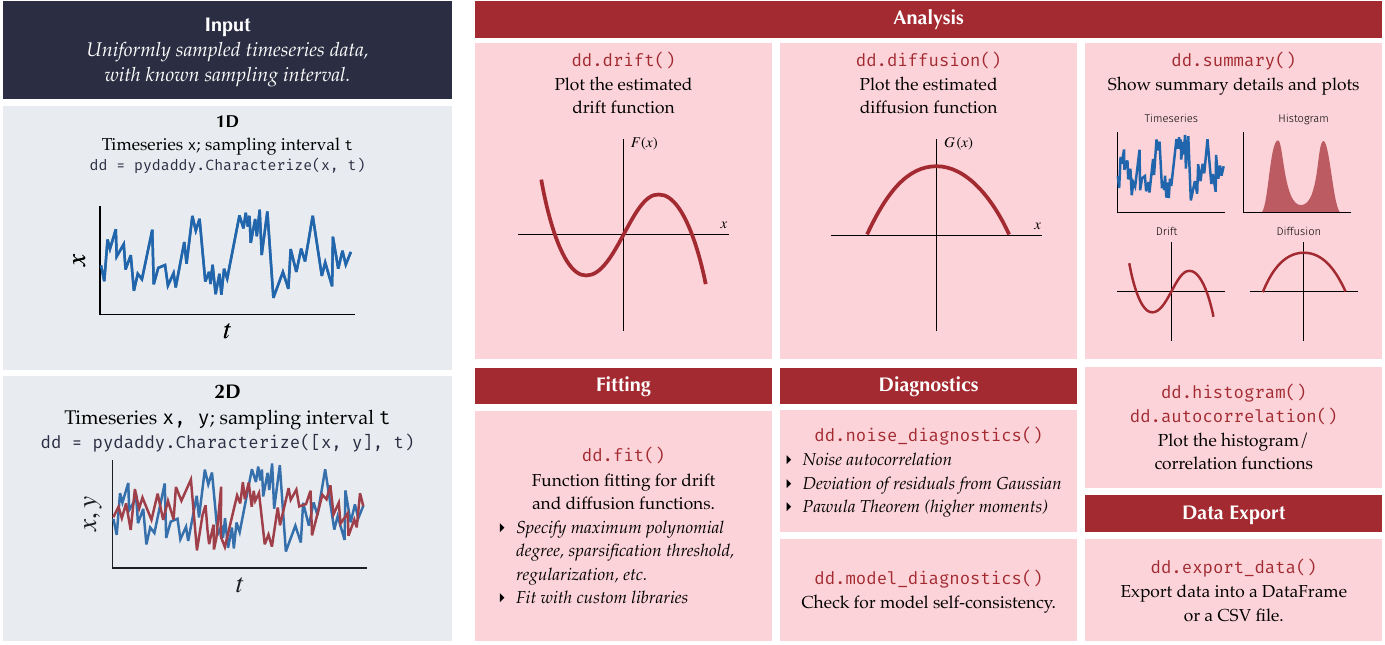}
    \caption{{\bf Overall schematics of the PyDaDDy package.}
PyDaDDy takes as input uniformly sampled 1D or 2D time series, and computes the drift and diffusion components from the time series. Several functions are provided to visualize data as time series or histograms, fit drift and diffusion functions, diagnose whether underlying assumptions for drift-diffusion estimation are met, and to export data.
    }
    \label{fig:schematics}
\end{figure*}

We make the above SDE estimation procedure available via an easy-to-use open-source python package named \textbf{PyDaDDy} which stands for {\bf Py}thon library for \textbf{Da}ta \textbf{D}riven \textbf{Dy}namics~\cite{pydaddy-zenodo}, available via \url{https://github.com/tee-lab/PyDaddy} (archived at \url{https://doi.org/10.5281/zenodo.13777396}~\cite{pydaddy-zenodo}). Figure~\ref{fig:schematics} shows an overview of the package. A brief walk-through of the package and its key features can be found in SI Section S2 -- \emph{PyDaDDy package -- a brief introduction}. Codes that implement the workflow are available as Jupyter notebooks as a part of our open-source package. The tool can be accessed even without installation via Google Colab: \url{https://pydaddy.readthedocs.io/en/latest/tutorials.html}. 

\section*{Results}
\subsection*{Discovering contrasting SDEs from synthetic datasets} 

\subsubsection*{Unimodal distributions from contrasting SDEs}

\begin{figure*}
    \includegraphics[width=\textwidth]{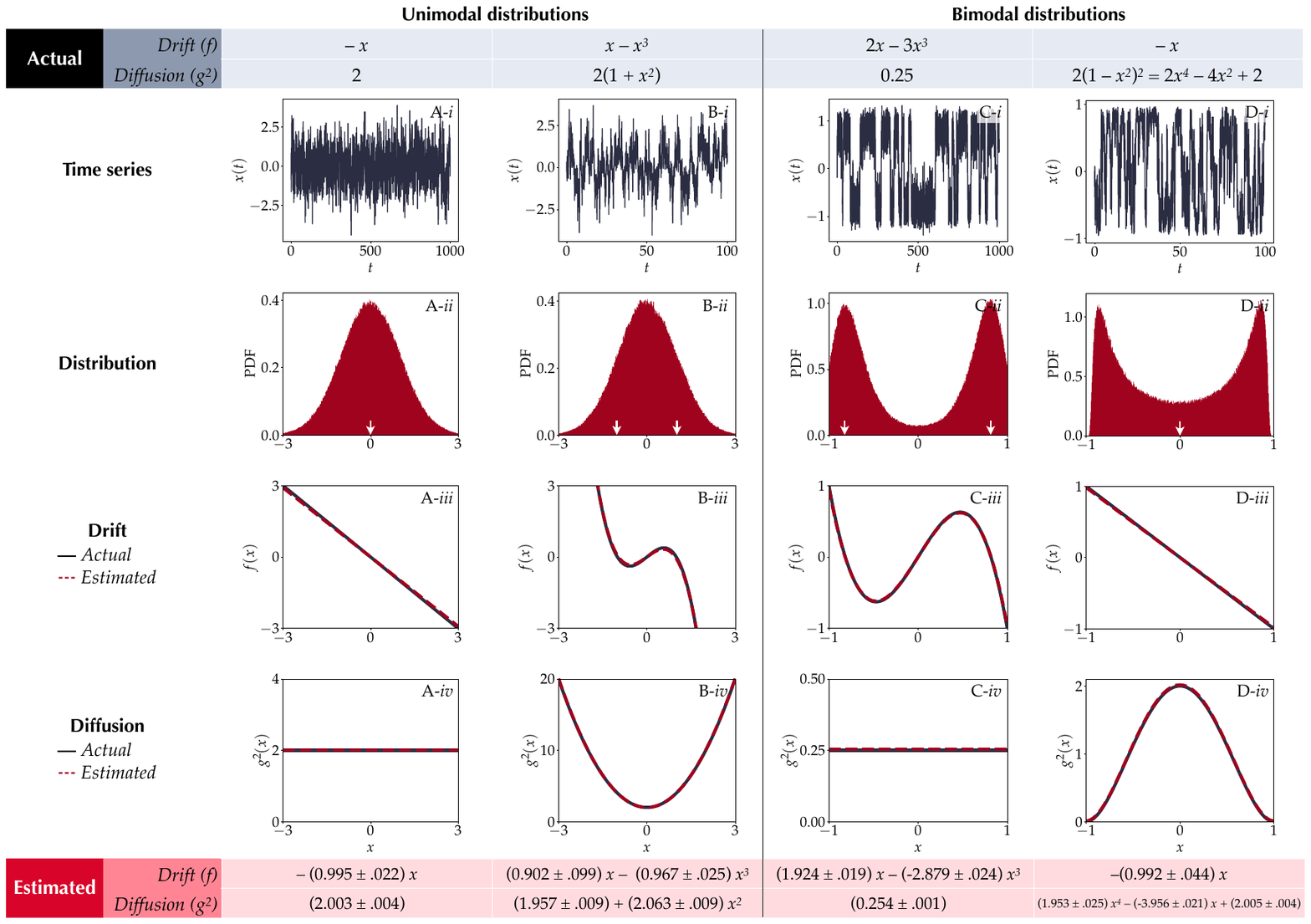}
    \caption{
    \textbf{Sparse regression on the extracted instantaneous jump moments reliably reconstructs governing equations from simulated data, for a wide range of dynamics.}
    The mathematical expressions for the actual and estimated drift and diffusion functions are shown in the shaded boxes on the top (grey-shaded) and bottom (red-shaded), respectively.
    Row \emph{i} shows a sample of the time series $x(t)$ of the models Eqs~\ref{eq:m1}-\ref{eq:m4}, respectively, across four columns  \emph{(A-D)},  while row \emph{ii} shows the corresponding histograms of $x(t)$. The arrows in histograms in row \emph{ii} mark the deterministic stable equilibria of the respective systems. Notice that the histograms in (A) and (B) are nearly identical, despite the underlying SDEs being very different. Similarly, the histograms in (C) and (D) are similar.
    \emph{(A)} A deterministic unimodal  system 
    (from Eq~\ref{eq:m1}.
    \emph{(B)} A noise-induced unimodal  state (from Eq.~\ref{eq:m2}).
    \emph{(C)} A deterministic bimodal system from Eq.~\ref{eq:m3}), with noise facilitating transitions between two deterministic states.
    \emph{(D)} A noise-induced bimodal  state (from Eq.~\ref{eq:m4}).
    Rows \emph{iii} and \emph{iv} compare the ground-truth (black) drift and diffusion functions with the estimated ones. The functions estimated using sparse regression are shown as dark red dashed lines. In all cases, the proposed method accurately recovers the drift and diffusion functions.
    \label{fig:synthetic}}
\end{figure*}

Unimodal distributions are widely observed in many biological datasets. However, they may be generated from very different dynamical processes and thus can be modeled by contrasting underlying SDEs. We aim to show that even in such cases, we can discover the original SDEs by applying our proposed method to the time series data. We consider the following two one-dimensional SDEs:

\begin{align}
    \dot x &= -x + \sqrt 2 \cdot \eta(t) \label{eq:m1} \\
    \dot x &= x - x^3 +  2 \left(\sqrt{1 + x^2}\right) \cdot \eta(t) \label{eq:m2}
\end{align}
Although Eq.~\ref{eq:m1} and \ref{eq:m2} are quite different, the time series generated with these equations, as well as their histograms, are very similar (Fig.~\ref{fig:synthetic} (A-i, A-ii, B-i, B-ii): in fact, the steady-state distributions of $x(t)$ are unimodal and are nearly identical for Eq.~\ref{eq:m1} and \ref{eq:m2}. From a dynamical systems point of view, Eq.~\ref{eq:m1} has a single deterministic stable state at $x^*=0$. Biologically, $x$ could be considered the deviation of the population size from its stable carrying capacity. Therefore, $x$ could be positive or negative. The additive noise---representing environmental fluctuations---simply spreads the dynamics of the population around the deterministic equilibrium, as reflected in the time series as well as the histogram (Fig.~\ref{fig:synthetic} (A-ii)). Therefore, we may refer to the mode of the histogram as reflecting a {\it deterministic state}. This is indeed the most commonly understood effect of noise.

On the other hand, for Eq~\ref{eq:m2} the deterministic stable equilibria are at $\pm 1$ but the mode of the histogram is at 0 (Fig.~\ref{fig:synthetic} (B-ii)). In this case, the state-dependent noise or the multiplicative noise-term completely alters the stability landscape, creating a new state between two deterministic stable states, leading to a mode at $x=0$. This is an example of an unusual effect of noise,  called the noise-induced stable state proposed in a dryland ecosystem model~\cite{odorico2005pnas}. 

For these unimodal datasets, we now pose the inverse problem: can we infer the correct underlying SDE models based on the features the time series data (Fig.~\ref{fig:synthetic} A-i versus B-i). Indeed, our approach integrating jump moment computations with sparse regression accurately identifies the functional forms of the drift and diffusion for both the models (in Fig.~\ref{fig:synthetic} compare red and black lines within columns A and B; rows iii and iv). Crucially, we are able to infer that the time series of Fig.~\ref{fig:synthetic} A-i is governed by a linear drift (Fig.~\ref{fig:synthetic} A-iii) with additive noise (Fig.~\ref{fig:synthetic} A-iv); in contrast, the time series of Fig.~\ref{fig:synthetic} B-i is driven by a nonlinear drift function with three roots (Fig.~\ref{fig:synthetic} B-iii) and a multiplicative noise (Fig.~\ref{fig:synthetic} B-iv). Furthermore, we are also able to recover the symbolic analytical expressions for the drift and diffusion functions which closely match the original SDEs that we deployed to generate the synthetic datasets (Fig.~\ref{fig:synthetic}, \emph{`Estimated'} panel).  

\subsubsection*{Bimodal distributions from contrasting SDEs}

Many biological systems exhibit alternative stable states, the simplest case being a bistable system, e.g. grassland and woodland states in dryland ecosystems or eutrophic and oligotrophic states of lakes. Systems with bistability show bimodal distributions of their state variable. Here too, bimodality can be generated by contrasting underlying processes/SDEs. To illustrate this, we consider the following two toy models:

\begin{align}
    \dot x &= 2x - 3x^3 + \frac12 \cdot \eta(t) \label{eq:m3} \\
    \dot x &= -x + \sqrt{2} (1 - x^2) \cdot \eta(t) \label{eq:m4}
\end{align}

Eq.~\ref{eq:m3} has two deterministic stable equilibria at $\pm \sqrt{2/3}$, and the dynamics is spread around these equilibria by the additive noise (Fig.~\ref{fig:synthetic} (C-ii)); thus, these are deterministic states. On the other hand, Eq.~\ref{eq:m4} has only one stable equilibrium at $x^*=0$; yet, due to the effect of the multiplicative noise term, the system exhibits two modes in the histogram away from the deterministic equilibrium (Fig.~\ref{fig:synthetic} (D-ii); thus, these are noise-induced states.

We again pose the inverse problem for these bimodal time series datasets (Fig.~\ref{fig:synthetic} C-i versus D-i and Fig.~\ref{fig:synthetic} C-ii versus D-ii). Here too, our method is able to recover the underlying SDE models, including the symbolic functions, correctly (see rows iii, iv, and\emph{`Estimated'} panels of Fig .~\ref{fig:synthetic} C and D). 


\subsection*{\label{sec:eco-models} Demonstration with classical models in theoretical ecology}
\begin{table*}
    \begin{tabular}{lll}
         \toprule
         \textbf{Model} & \textbf{Equations} & \textbf{Estimated Equations} \\         
         \midrule
         
         Logistic model      & 
         $\begin{aligned}
            f(N) &= 2N - 0.5N^2 \\
            g^2(N) &= N
         \end{aligned}$   &
         $\begin{aligned}
            f(N) &= 2.03N - 0.41N^2 \\
            g^2(N) &= 1.07N
         \end{aligned}$    \\
         \midrule

         Harvesting model    &
         $\begin{aligned}
             f(N) &= 2.4N - 0.4N^2 - 4\frac{N^2}{1+N^2} \\
             g^2(N) &= 0.04 \left( \frac{N^2}{1 + N^2}\right)^2
         \end{aligned}$ &
         $\begin{aligned}
             f(N) &= 2.77N - 0.46N^2 - 4.61 \frac{N^2}{1+N^2} \\
             g^2(N) &= 0.04 \left( \frac{N^2}{1 + N^2}\right)^2
         \end{aligned}$ \\
         \midrule

         Lake eutrophication model     &
         $\begin{aligned}
             f(x) &= 0.5 - x + \frac{x^8}{1 + x^8} \\
             g^2(x) &= 0.04
         \end{aligned}$ &
         $\begin{aligned}
             f(x) &= 0.53 - 1.05x + 1.06\frac{x^8}{1 + x^8} \\
             g^2(x) &= 0.04
         \end{aligned}$ \\
         \midrule

         Lotka-Volterra competition     &
         $\begin{aligned}
             f_1(x, y) &= 2x - 0.3x^2 + 0.3xy \\
             f_2(x, y) &= 4y - 0.5y^2 + xy \\
             g_{11}^2(x, y) &= 0.04x^2 \\
             g_{22}^2(x, y) &= 0.04y^2
         \end{aligned}$ & 
         $\begin{aligned}
             f_1(x, y) &= 2.14x - 0.36x^2 + .38xy \\
             f_2(x, y) &= 3.99y - 0.49y^2 + 1.02xy \\
             g_{11}^2(x, y) &= 0.04x^2 \\
             g_{22}^2(x, y) &= 0.04y^2
         \end{aligned}$ \\
        \midrule
        
         Van der Pol &
         $\begin{aligned}
             f_1(x, v) &= v \\
             f_2(x, v) &= -x + 5v - 5x^2v\\
             g_{22}^2(x, v) &= 25
         \end{aligned}$ &
         $\begin{aligned}
             f_1(x, v) &= v \\
             f_2(x, v) &= -1.04x + 4.95v - 4.99x^2v\\
             g_{22}^2(x, v) &= 25.61
         \end{aligned}$ 
         \\ \midrule

         Prey Predator &
         $\begin{aligned}
             f_1(n, p) &= n-n^2-1.2\frac{np}{n+p} \\
             f_2(n, p) &= 0.48\frac{np}{n+p}-0.2p\\
             g_{11}^2(n, p) &= 0.01n^2 \\
             g_{22}^2(n, p) &= 0.01p^2
         \end{aligned}$ &
         $\begin{aligned}
             f_1(n, p) &= 1.02n-1.03n^2-1.22\frac{np}{n+p} \\
             f_2(n, p) &= 0.49\frac{np}{n+p}-0.20p\\
             g_{11}^2(n, p) &= 0.01n^2 \\
             g_{22}^2(n, p) &= 0.01p^2
         \end{aligned}$ 
         \\ \bottomrule         
    \end{tabular}
    
    \caption{\textbf{Reconstructing classical models from the analysis of time series fluctuations.} Comparison of reconstructed SDEs with original SDEs for several classical models in ecology. Each model is an SDE of the form $\dot x = f(x) + g(x) \cdot \eta(t)$. For the vector models, the drift $f$ had components $f_1(x, y)$ and $f_2(x, y)$, and diffusion had components $g_{11}$, $g_{22}$ (the off-diagonal, cross-diffusion components $g_{12} = g_{21}$ were 0 for the models considered.}
    \label{tab:models}
\end{table*}

We now validate the method against several classical models in theoretical ecology. We considered stochastic modifications in various models, with one and two variables. The following univariate models were considered: the logistic model of density-dependent population growth (unistable system), a population harvesting model with a Holling's type III functional response (a bistable system)~\cite{strogatz2018nonlinear} and a lake eutrophication model~\cite{carpenter1999management} (a bistable system). The bivariate models considered were: the Lotka-Volterra competition model for inter-specific competition, a non-linear predator-prey model with a Holling's type II functional response for predation~\cite{alonso2002mutual}, and the Van der Pol oscillator, a minimal model for non-linear oscillations~\cite{strogatz2018nonlinear}. 

Table~\ref{tab:models} summarizes the results of this analysis. In summary, for all the models considered, the SDE discovery procedure was able to accurately discover the correct model from the simulated time series (see SI Section S3 -- \emph{Demonstration with classic models in ecology} for more details).

\subsubsection*{Inaccurate model discovery with limited data}
Real-world data has many limitations, such as short time series not necessarily sampled at high enough resolution. It is important to understand how such contexts may affect the data-discovery protocol we described. Here, we examine two specific sub-cases of \emph{`limited data'}: short time series, and long sampling interval. 

To do so, we attempt to estimate the model in Eqn.~4 described in the previous section, $\dot x = 2x - 3x^3 + \frac12 \eta(t)$, which exhibits bistable dynamics. We then use a short time series (1000 high-resolution observations, with $\Delta t = 0.01 s$) where the system has explored only a limited part of the state-space. The resultant model, along with the results of the model diagnostics, are shown in SI Fig~S11 A, B. Not surprisingly, the estimated SDE does not match the true ground-truth model. A similar mismatch happens when one estimates a model from data observed with a large sampling interval (SI Fig~S11 C, D). Here, the large sampling interval means that information about the fine-scale dynamics is lost, so the discovered model again deviates from the actual model. However, in both cases, we observe a large mismatch between the autocorrelation functions of the original and model-simulated time series, which will alert the user that the model estimation is inaccurate. Another tell-tale sign of a potentially misidentified model, due to limited or noisy data, is large error bars in the estimated coefficients, which we also observe. We extensively address the issues of limitations, avoiding pitfalls while using the methods, evaluating the discovered model with diagnostics in the Discussion section and quantitative characterization of the limitations in SI Section S5 - \emph{Pitfalls and limitations}.  

\subsection*{Application to real datasets}
At the outset, it is not obvious that these methods are applicable to complex biological datasets. In particular, SDEs assume a Gaussian uncorrelated noise, while we often expect more complicated noise structures (e.g., non-Gaussian and with memory) for real datasets. Therefore, we demonstrate the applicability and generality by using them on two contrasting datasets, a dataset of group trajectories of schooling fish~\cite{jhawar2020fish}, and another of confined single-cell migration~\cite{bruckner2019stochastic}. 

The two datasets are different in many important ways. First, the fish schooling dataset captures emergent group-level order, while the cell dataset is an individual-level behavior. Second, the fish school has a relatively spatial span ($\sim 1$ m) but has relatively fast temporal dynamics ($\sim 0.1$ s). In contrast, the cell dataset has a small spatial span ($\sim 10^{-4}$ m), but the cells move very slowly, with a timescale of $~10$ min. Finally, their underlying dynamics are fundamentally different---while fish schools are in a noise-induced state~\cite{jhawar2020fish}, the cell movement is predominantly a deterministic limit cycle with noise playing a minor role~\cite{bruckner2019stochastic}.

\subsubsection*{SDE discovery of noise-induced schooling of fish}
Among various fields of biology, collective animal motion has advanced substantially over the last decade. Employing state-of-the-art imaging technologies to record moving collectives,  researchers have produced highly resolved spatiotemporal datasets that help us probe into the mechanisms of synchronized collective motion. Mathematical theories of flocking express the dynamics of collective motion in the form of stochastic differential equations---offering potential explanations for such synchronous group dynamics~\cite{vicsek2012collective,toner2005hydrodynamics,ramaswamy2017active,biancalani2014prl,jhawar2019bookchapter}. With the availability of high-quality animal movement data, we now pose the inverse problem, i.e. can a stochastic dynamical model be discovered from given time series of animal trajectories?  Further, one can address fundamental questions on the role of stochasticity in shaping (or destroying) the order. For example, is the observed collective dynamics consistent with a deterministic state (i.e. does stochasticity merely blur the order around a deterministic stable equilibrium as in Fig~\ref{fig:synthetic}(A-\textit{ii})) or a noise-induced state (i.e. does stochasticity create nontrivial states away from the deterministic stable equilibrium, as in Fig~\ref{fig:synthetic}(C-\textit{ii}))?

A recent study inferred that the highly synchronized motion of schooling fish is a noise-induced state~\cite{jhawar2020fish}. To do so, they employed the conventional jump moment estimation on the time series data of group dynamics. Here, we test if we can recover the same results using our inference protocol that integrates jump moment computation with the sparse regression. Further, we perform noise and model diagnostics, which were ignored in that study. 

We use the openly available published dataset of a group of \emph{Etroplus suratensis} (karimeen), from~\cite{jhawar2020fish}. Here, the motion of a school of 15 fish was recorded using a high-resolution camera and was subsequently tracked using computer vision methods. This is a 30-dimensional time series, consisting of the 2D positions of each of the 15 fish at each point in time. Trying to model the dynamics in this 30-dimensional space is perhaps not very informative--- low-dimensional representations of the trajectory, capturing the essential dynamics of the group, are more suitable for gaining insights into the collective dynamics. Specifically, motivated by physics literature~\cite{vicsek1995novel,toner2005hydrodynamics} and its applications in many biological collective motion~\cite{couzin2002collective,tunstrom2013collective,jhawar2020fish}, we would like to study how polarization order emerges and evolves in the group dynamics.

Thus, we use a two-dimensional \emph{group polarization} vector, denoted $\mathbf m = (m_x, m_y)$: This is simply the vector average of the directions of all fish, with the following interpretation. If the school of fish is highly ordered, with nearly all fish moving in the same direction, the magnitude of the polarization vector, $|\mathbf m|$, will be close to 1. On the other hand, if the group is disordered, with fish moving in random directions, $|\mathbf m|$, will be close to 0. The polarization order parameter is widely used in biology as well as physics literature to characterize the degree of order, in terms of synchronized motion, in collective movement literature~\cite{vicsek1995novel,chate2020dadam,couzin2002collective,tunstrom2013collective} and summarizes the state of the entire group as a single vector. Studying the stochastic dynamics of polarization will help us gain insights into the nature of order in the system.

\begin{figure*}
    \includegraphics[width=\textwidth]{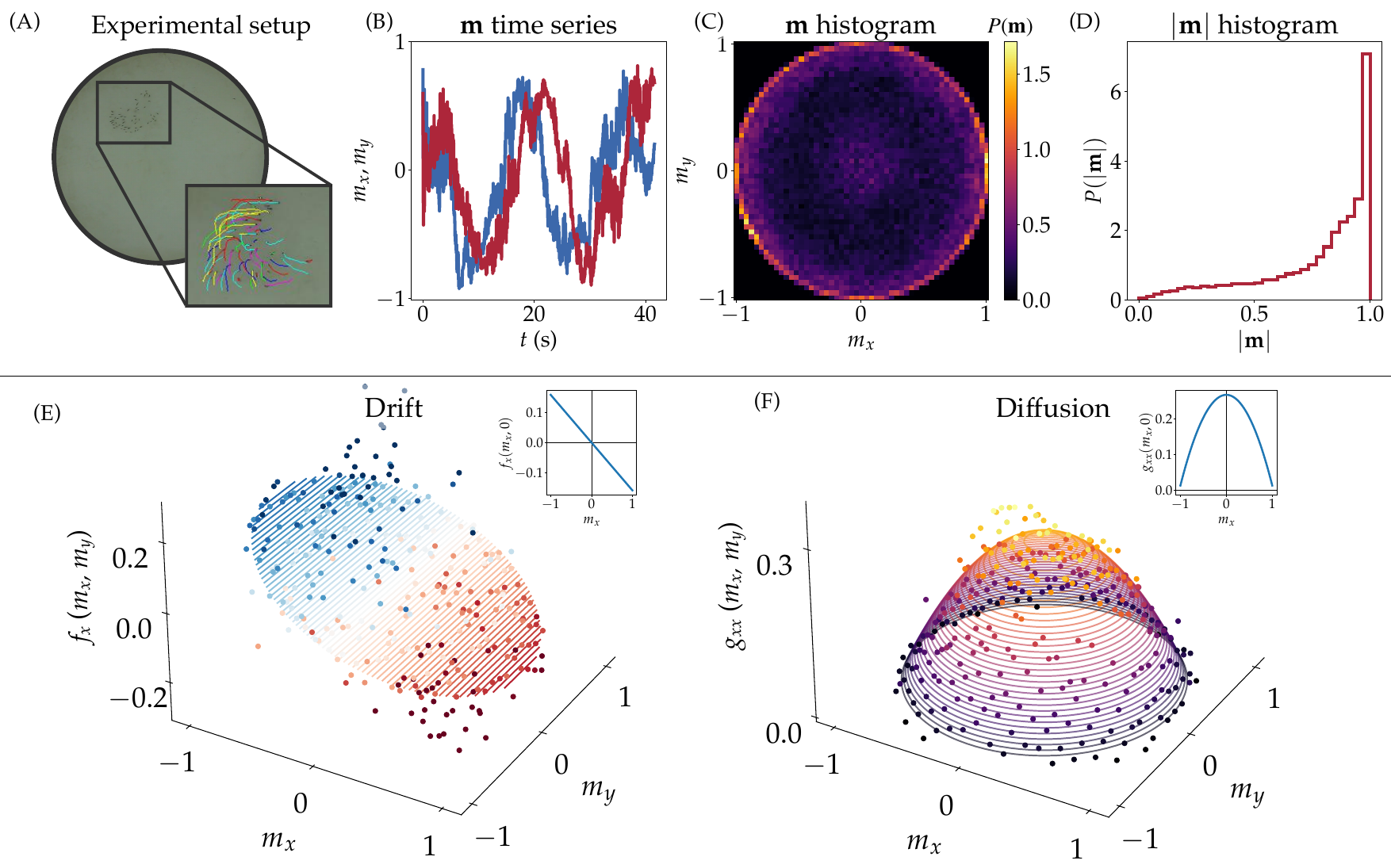}
    \caption{\textbf{Data driven SDE discovery for polarization dynamics in fish schools; data from~\cite{jhawar2020fish}.} (A) Experimental setup and individual fish trajectories (in different colours) are extracted from video recordings of fish swimming in a tank (from~\cite{jhawar2020fish}.). (B) From the individual trajectories, a time series of the polarization vector, $\bm$, is computed. The trajectories shown are the $x$- (red) and $y$- (blue) components of $\bm$. (C, D) Histograms of the polarization vector $\bm$ and the net polarization, $\modm$, respectively. (E) The drift function discovered by the SDE discovery procedure. The $x$-component of the function, $f_x$ is shown. The surface plot shows the fitted drift function, and the points show the binwise averaged estimates. Inset shows a slice of this function along the $y=0$ plane, showing a single stable equilibrium at $x = 0$ (F) The discovered diffusion function. The inset shows a slice along the $y=0$ plane. The diffusion is maximum at $\bm = 0$, and decreases outwards. In (E) and (F), the colours redundantly represent $f_x$ and $G_{xx}$, respectively and are added only for visual clarity.}
    \label{fig:fish-schooling-data}
\end{figure*}

The dataset contains the time series of $\mathbf m$, available at a uniform interval of 0.12 seconds for roughly one hour, with several missing data points (corresponding to time points where the tracking was imperfect). Although the time series of the group polarization $\mathbf m$ shows significant stochastic fluctuations (Fig \ref{fig:fish-schooling-data}B),  the histograms show that the fish school is predominantly ordered (Fig 3C, D). This forms the input time series dataset for our analysis. 

Here, we can recover the same results by using our protocol---implemented via the package PyDaDDy---that combines jump moment estimation and sparse regression (also see SI Section S4 A -- \emph{Model selection for the fish schooling dataset}). Our discovered equation contains a linear drift and a quadratic diffusion, with a vector SDE  of the form:
\begin{align}
\dot {\mathbf m} = - \alpha \mathbf m + \sqrt{\beta (1 - | \mathbf m |^2)} \cdot \boldsymbol \eta(t),
\label{eqn:fish-eq-inferred}
\end{align}
with $\alpha = 0.16$ and $\beta = 0.26$.

The deterministic stable equilibrium of this data-discovered SDE is a disordered state, i.e. ${\mathbf m^*} = 0$ (Fig \ref{fig:fish-schooling-data}E). However, the observed state, or the mode of the order parameter, is at ${|\mathbf m| \approx 1}$, away from the deterministic stable state. This leads to the counter-intuitive implication that schooling fish is a noise-induced state (we refer the readers to~\cite{jhawar2020fish} for further discussion on this intriguing phenomenon).

\begin{figure*}
    \includegraphics[width=\textwidth]{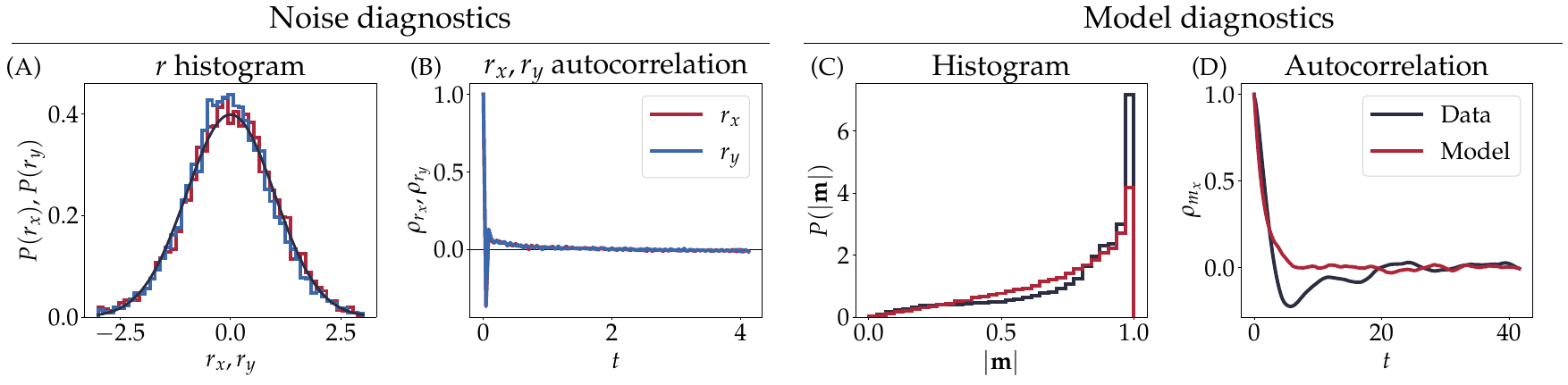}
    \caption{{\bf Diagnostics of the discovered SDE for fish polarization dynamics.}
    (A, B) Noise diagnostics. (A) The histogram of the residual noise $r$ is computed based on the discovered SDE. The $r_x$ and $r_y$ marginals are shown as red and blue. These should match a standard normal distribution, shown in black. (B) The autocorrelation functions of $r_x$ and $r_y$. The autocorrelation should decay within one sampling time step.
    (C, D) Model diagnostics.
    (C) Histogram of $\modm$ from the original time series (black) compared to that of a simulated time series generated from the discovered SDE (red).
    (D) Autocorrelation of the $m_x$ component of the real (black) and simulated (red) time series.
    }
    \label{fig:fish-schooling-diag}
\end{figure*}

Fig.~\ref{fig:fish-schooling-diag} shows the results from diagnostic tests on the discovered SDE model. The noise residual $r(t)$ has a Gaussian distribution, as expected (Fig.~\ref{fig:fish-schooling-diag}A). The correlations in the residual noise $r(t)$ decay rapidly (Fig.~\ref{fig:fish-schooling-diag}B). These tests reasonably support the modelling assumption of $\eta$ being Gaussian white noise.

We also find that the discovered equations pass model diagnostic tests. The histogram of simulated SDE from Eq~\ref{eqn:fish-eq-inferred} closely matches the histogram of $\bm$ from the original time series~\ref{fig:fish-schooling-diag}C).  The autocorrelation of the simulated time series also shows reasonable agreement with the original time series~\ref{fig:fish-schooling-diag}D), with one notable deviation: The data autocorrelation shows negative values over relatively larger time scales of $\approx$ 10 s, before converging to zero. The authors of the fish schooling paper demonstrated that this feature arises from boundary effects that are unimportant for schooling dynamics of {\it Etroplus suratensis}. Nevertheless, we emphasize that the SDE discovery protocol used only highly local information on group polarization fluctuations at very small time scales $\approx 0.1 s$; it did not use any information on the frequency distribution or the autocorrelation function of the group polarization data. Despite this, the simulated time series shows a good agreement with data in both these metrics. Last, but not the least, we find that the model is self-consistent, with the same SDEs being recovered when we pass the simulated data via our equation discovery protocol. 

\subsubsection*{SDE discovery for dynamics of confined cell migration}
\begin{figure*}
    \includegraphics[width=\textwidth]{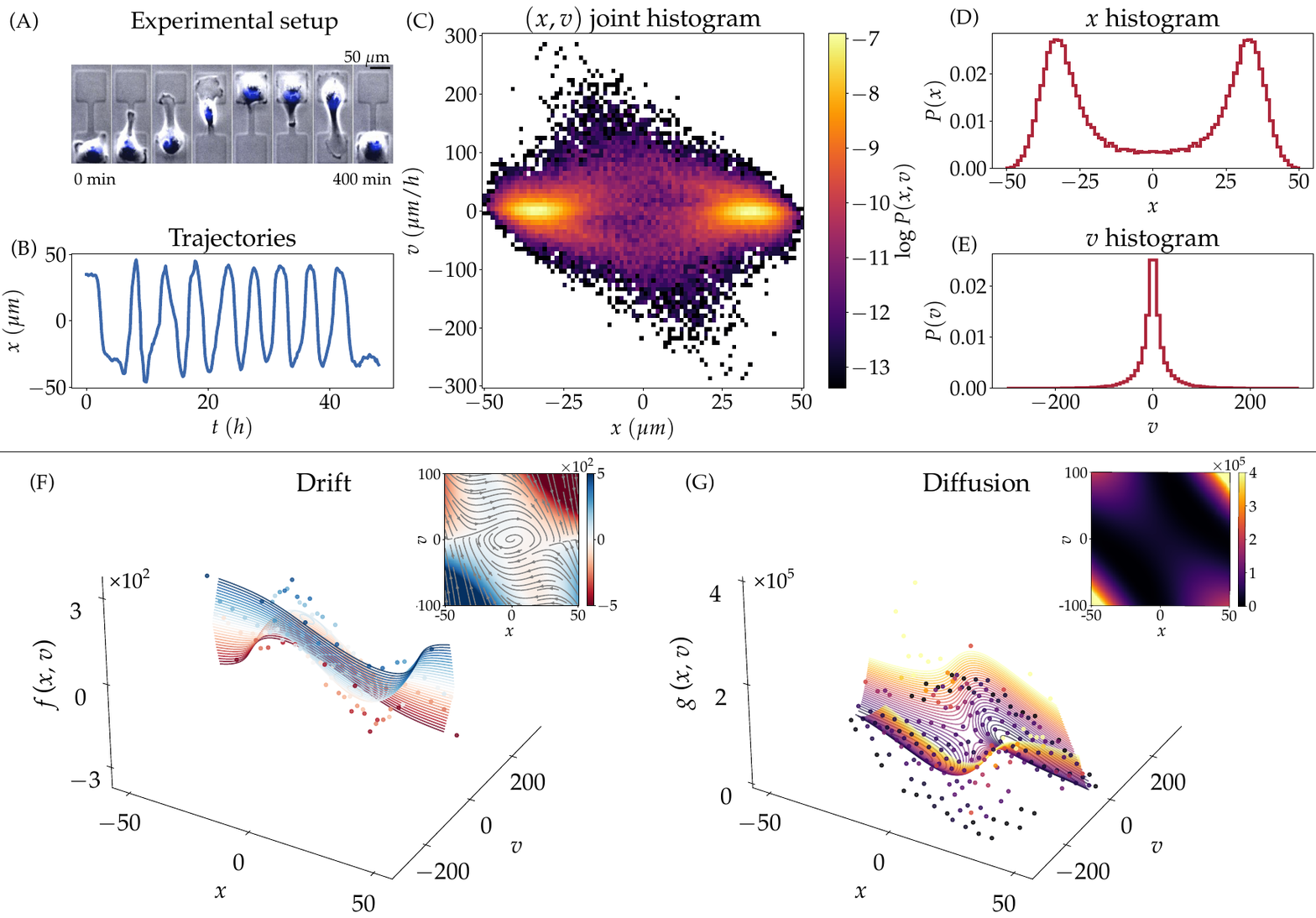}
    \caption{{\bf Discovering SDEs governing the dynamics of confined migration of cancer cells; data from~\cite{bruckner2019stochastic}}
    (A) The experimental setup, showing a micropattern with two landing pads, and the cell hopping between them~(from~\cite{bruckner2019stochastic}). 
    (B) Example time series of cell position. (C) Joint histogram of the position $x$ and velocity $v$ of the cell trajectories. (D, E) Marginal histograms of $x$ and $v$.
    (F, G) The discovered SDE for the dynamics of $v$. (F) The drift function, $f(x, v)$. The surface plot shows the fitted drift function, and the points show the binwise averaged estimates. The inset shows the direction field of the deterministic dynamics, with the background colour denoting $f$.
    (G) The diffusion function, $g(x, v)$. The inset shows the diffusion as a colour map. Since $g$ is relatively higher near the limit cycle, the actual observed cycles have slightly larger amplitude. In (F) and (G), the colours redundantly represent $f_x$ and $G_{xx}$, respectively and are added only for visual clarity.
    }
    \label{fig:cell-mig-data}
\end{figure*}

From morphogenesis and wound healing to cancer metastasis, cell migration plays a key role in many biological contexts, where it is crucial to understand how cells move through complex environments. A recent study~ \cite{bruckner2019stochastic} explored the stochastic dynamics of cell migration in structured environments. The authors designed an experiment where a cancerous cell migrates back and forth between two \emph{islands} (`states')~(Fig~\ref{fig:cell-mig-data}A), and modeled the dynamics of this two-state migration using an SDE framework, but without using an explicit expression. They concluded that the dynamics is primarily governed by a limit cycle in the deterministic (drift) component. The stochastic component plays a relatively minor role affecting only the timescale of the transition between states. Here, we reproduce the key results from this study using our SDE discovery protocol. While the original study represented the SDE using bin-averaged, piecewise representations of the drift and diffusion functions, our methodology enables us to extract interpretable functional forms. Further, we demonstrate how the noise and model diagnostics can lead to a more accurate model discovery. 

The dataset consists of 149 independent replicate cell trajectories. Each trajectory was based on high-resolution images taken once every 10 minutes, for up to 50 hours. Fig.~\ref{fig:cell-mig-data}B shows an example time series of the cell movement between the two islands and Fig.~\ref{fig:cell-mig-data}C-E depict histograms of the state variables (position $x$ and velocity $v$).

Based on the experimental design (Fig.~\ref{fig:cell-mig-data}A) as well as the bimodal histograms of cell positions (Fig.~\ref{fig:cell-mig-data}C), a natural first attempt would be to model the cell hopping between two stable states, viz. the landing pads, with the switching being purely stochastic. This is analogous to the model described in Fig~\ref{fig:synthetic}C, which exhibits stochastic transitions between two stable states. This would involve modelling the cell trajectory $x(t)$ as an \emph{overdamped} SDE of the following form:

\begin{align}
    \dot x = f(x) + g(x) \cdot \eta(t)
\end{align}

However, our model diagnostics reveal that the overdamped model is not sufficient to fully capture the dynamics of the system (see SI Section S4 B -- \emph{Model selection for the cell migration dataset}). Thus, we model the cell trajectory using an \emph{underdamped} model with two dynamical variables, position ($x$) and velocity ($v$). Such a two-dimensional description would be able to capture, for instance, linear or non-linear oscillatory behaviour in the system; this indeed best describes the data, as we show below. The underdamped stochastic dynamics follows

\begin{align}
    \dot x &= v \\
    \dot v &= f(x, v) + g(x, v) \cdot \eta(t)
\end{align}
where we scale $x$ and $v$ to non-dimensionalize them before applying the SDE discovery procedure. This scaling step is necessary for the sparse regression to work correctly (see SI Section S4 B). Using our SDE discovery procedure, we find that the drift function $f$ (Fig.~\ref{fig:cell-mig-data}E) is well approximated by a cubic polynomial of the form:
\begin{align}
    f(x, v) = \underbrace{(\alpha_1 - \alpha_2 x^2)v - \alpha_3 x}_{\text{Terms in Van der pol equation}}  + \underbrace{\alpha_4 x^3 - \alpha_5 v^3 - \alpha_6 xv^2}_{\text{additional terms}} ,
    \label{eqn:cell-eqn-inferred}
\end{align}
\newcommand{\E}[1]{\times 10^{#1}}
with $\alpha_1 = 49.84, \alpha_2 = -3.50 \E{-3}, \alpha_3 = -1.27, \alpha_4 = 5.51 \E{-4}, \alpha_5 = -2.01 \E{-4}$, and $\alpha_6 = -1.49 \E{-3}$.

Further, we discover a multiplicative noise for the diffusion function $g^2(x, v)$, approximated to a 4th order polynomial of the form, (Fig.~\ref{fig:cell-mig-data}F).\begin{align}
    g^2(x, v) = & \; \beta_1 - \beta_2 x^2 + \beta_3 x^4 - \beta_4 xv \nonumber \\
              & + \beta_5 x^3v - \beta_6 v^2 + \beta_7 x^2v^2 + \beta_8 xv^3 + \beta_9 v^4,
              \label{eqn:cell-eqn-inferred-g}
\end{align}

with $\beta_1 = 3.20 \E{4}, \beta_2 = -1.20 \E{-2}, \beta_3 = 1.25 \E{-9}, \beta_4 = -3.34 \E{-3}, \beta_5 = 6.53 \E{-10}, \beta_6 = -5.16 \E{-5}, \beta_7 = 8.13 \E{-11}, \beta_8 = 2.62 \E{-12}$ and $\beta_9 = -2.87 \E{-14}$.

The actual values of the coefficients, in both the scaled and unscaled coordinates, are shown in SI Tables~S1 and S2.

Our data-discovered SDE model of cell migration yields an excitable flow, qualitatively consistent with the findings of~\cite{bruckner2019stochastic}, exhibiting relaxation oscillations. Furthermore, we see that the drift function deviates from the classic Van der Pol oscillator, which is a minimal model that explains relaxation oscillations.
Intriguingly, as shown in the SI Section S4 B, we find that this diffusion term is necessary to capture the finite boundary effects on the dynamics of cells and to satisfy model diagnostics (see below).

\begin{figure*}
    \includegraphics[width=\textwidth]{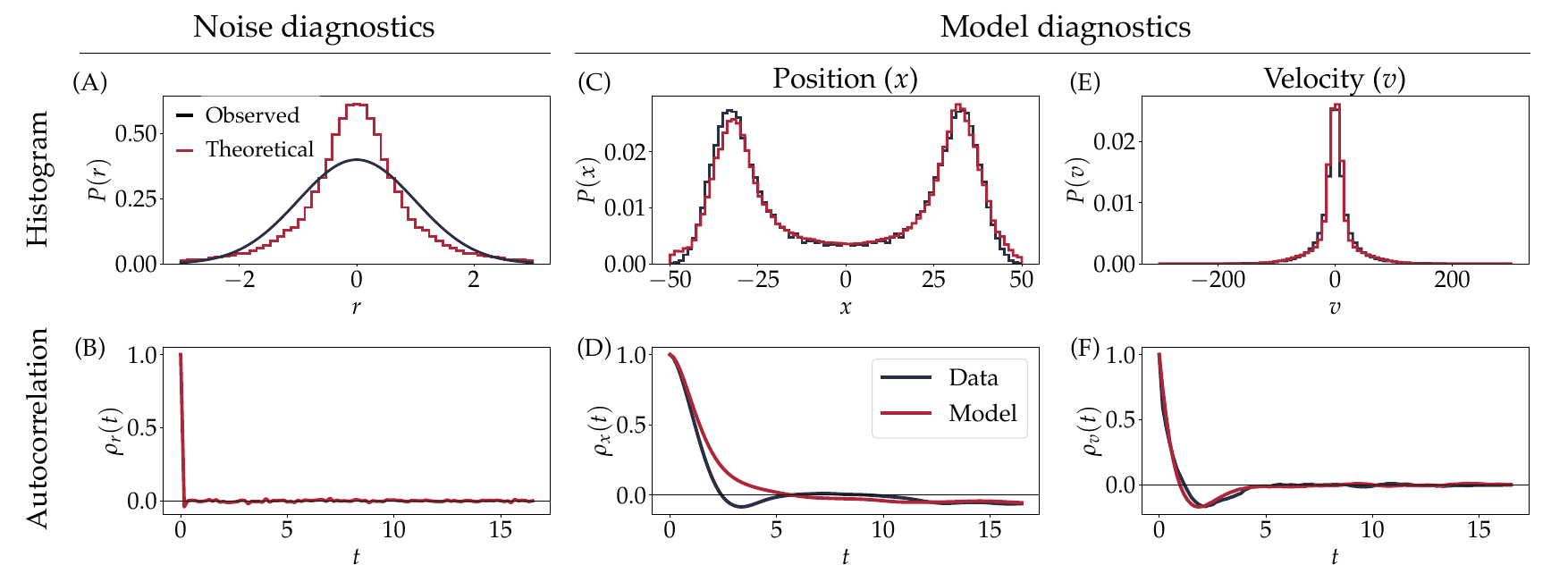}
    \caption{{\bf Diagnostics of the discovered SDE of cell migration.}
    (A, B) Noise diagnostics. (A) The histogram of the residual noise $r$, computed based on the discovered SDE and data. This should match a standard normal distribution, shown in black. (B) The autocorrelation function of the residual $r$.
    (C-F) Model diagnostics. (C, E) Histograms of the position and velocity of the original data (black) versus a time series simulated using the discovered model (red). (D, F) Autocorrelation functions of the position and velocity of the original data (black) versus a time series simulated using the discovered model (red). 
    }
    \label{fig:cell-mig-diag}
\end{figure*}

Fig.~\ref{fig:cell-mig-diag} shows the diagnostic results of the discovered model. The noise residuals show no correlation in time (\ref{fig:cell-mig-diag}B), but the distribution of the noise has a stronger central tendency than a Gaussian distribution (\ref{fig:cell-mig-diag}A). Nevertheless, the discovered model can produce simulated time series with statistical properties, i.e. state variable histograms and the autocorrelation functions, broadly consistent with the original data (\ref{fig:cell-mig-diag}C-F). Additionally, the model we have presented is self-consistent. Here, we highlight the important role played by model diagnostics in accurately discovering the diffusion term. An additive noise term (i.e. a constant diffusion) with the same drift function could not accurately capture the spatial constraints of the cell migration, with the simulated trajectories frequently crossing the locations beyond the boundary. This led to huge discrepancies between the histogram of the state variables of data and the simulated model. However, a fourth-order diffusion term minimised such physical barrier violations; this enabled better model self-consistency. Therefore, we interpret that the diffusion term effectively captures spatial constraints. 

\subsubsection*{Generalizability of the discovered models}

To confirm that the SDE discovery approach discovers generalizable models, we validate the discovered with left-out data (see subsection \emph{Diagnostics} -- Validation with left-out data) for both the fish and cell datasets. Here, we divide the datasets into two halves, a \emph{training set} and a \emph{validation set}. For the fish schooling dataset, two out of four available trials is used as the training set. For the cell migration dataset, 75 out of the 149 cell trajectories are used. We perform the model discovery procedure on the training set. We then compute the histogram and autocorrelation for a time series generated by the discovered model and the left-out data. For both the fish schooling and the cell migration dataset, the statistics of the model-generated time series match those of the validation data, suggesting that the discovered SDEs have captured generalizable features of the underlying dynamics (SI Section S4 C -- \emph{Generalizability of the discovered SDE models}).

\section*{Discussion}

In this paper, we have presented a method that takes empirically observed time series as the input and discovers a data-driven stochastic dynamical equation in an analytically interpretable form. To achieve this, we combined the traditional approaches of drift and diffusion estimation using jump moments~\cite{van1992stochastic,tabar2019book}, with techniques based on sparse regression~\cite{brunton2016sindy,boninsegna2018sparse}. Importantly, we emphasized the need for diagnostics to test the assumptions associated with data-derived models and the validity of the discovered model. In addition to illustrating the method in synthetic datasets where the ground truth equations are known, we demonstrated the generality of the method with applications to two contrasting biological time series datasets---one of schooling fish and the other of cell migration. To enable the use of this method to a larger audience of biological researchers, we implemented an open and easy-to-use Python package, PyDaDDy~\url {https://pydaddy.readthedocs.io/}.

We explain how this method can offer unique insights into the underlying dynamics of biological systems. Let us eye-ball the data of the group polarization variables ($m_x, m_y$ from Fig~\ref{fig:fish-schooling-data} A) from the schooling fish experiment and the cell migration data ($x, v$ from Fig~\ref{fig:cell-mig-data} A). Based on this, one may naively infer that they both seem to exhibit oscillatory patterns in their respective state variables, perhaps with different degrees of stochasticity. Some classic dynamical system based explanations for oscillations are that of a simple harmonic oscillator and a van der Pol oscillator, both of which are deterministic processes. One can also hypothesize more complex dynamical models/processes that offer explanations of the qualitative and quantitative features of the data. However, based on the data-driven equation discovery, we discovered a stochastic dynamical equation which reveals that schooling is a noise-induced state formed away from the deterministic stable equilibrium. On the other hand, the oscillatory patterns of the cell data were revealed to be due to a deterministic limit cycle, with noise playing a relatively minor role. Furthermore, the mathematical structure of this limit cycle is different from that of the classic van der Pol oscillator. In other words, we could eliminate several alternative hypotheses while arriving at data-driven dynamical models with symbolic representations with relatively minimal user inputs.


We highlight that the method is highly efficient computationally. For example, for the fish schooling dataset provided with the package (a 2D vector time series with around 25,000 time points), the initial estimation of the drift and diffusion coefficients takes approximately 800 milliseconds (ms) and the function fitting takes 2ms for the drift function and 6 ms for the diffusion function, on a 2023 Mac Mini (Apple M2 processor, 16GB RAM). On a Google Colab instance, these operations take around 2.5 s, 10 ms and 60 ms respectively.

\subsection*{Pitfalls and how to avoid them}

Real-world biological systems are complex, and the ``true" underlying dynamics may involve dozens if not hundreds of interacting variables. Further, observed data from such systems can be partial (with only a subset of the dynamical variables observed), imperfect (with measurement noise, or with data points few and far in between), incomplete (with data missing at several time points) or not fully representative of the full system dynamics (with data covering the full state space only partially).
It is not immediately clear that such systems can be modeled using stochastic differential equations involving only a few interacting variables. This poses the important question: how can one one ensure that a data-driven SDE derived from a given empirical dataset is indeed a valid model of the dynamics? If one blindly applies the SDE discovery pipeline to any dataset, the pipeline would produce some SDE. How can we verify that the returned SDE is indeed valid, and is a good representation of the system's dynamics? Hence, it is important to carefully evaluate the suitability of SDE discovery techniques to the specific problem or dataset being considered. Therefore, a key philosophy of our methodology (and the associated package PyDaDDy) is to equip the user with tools to evaluate and diagnose the discovered models and make informed decisions. In this section, and in SI Section S5, we discuss some pitfalls when trying to model biological systems using low-dimensional SDEs, ways to identify them and, when possible, fix them.

\paragraph{Dimensionality of the system} The data-driven equation discovery approach, in principle, works for any arbitrary dimension. In this manuscript, we chose to focus on one and two-dimensional datasets to elucidate the key principles and applications of inferring stochastic dynamical equations. To illustrate the generality of the method, we provide an example code for the recovery of the Lorenz equation (with stochasticity), a classic three-dimensional nonlinear model that exhibits chaos (see notebook:  \emph{Higher dimensions}). Some practical considerations while dealing with data-driven SDE models in high dimensions are worth noting. First, as the number of dimensions increases, the number of candidate functions in the library increases, and the estimation procedure can become error-prone. The sparse regression algorithm is a form of regularization which can counter this effect to an extent. In addition, knowledge about the physics, constraints and symmetries of the problem can sometimes help in cleverly designing a library with a small number of terms~\cite{nabeel2023data}. Moreover, validating the discovered models becomes a non-trivial problem in higher dimensions: quantitative and statistical techniques (without the benefits of the visual techniques feasible for 1 and 2-dim) will need to be appropriately adopted, a topic for future research.

We remark, however, that when the goal is to construct interpretable equations in terms of the dynamical variables, systems which are amenable to lower dimensional representations are particularly well suited for applying data-discovery procedures. For example, even when the observed dataset is high-dimensional (for example, the fish dataset is $2N$-dimensional, where $N$ is the number of fish), it is often possible to find a small number of latent dimensions which describe the essential dynamics of the system. Often, such a low-dimensional representation is not merely convenient for analysis but is also desirable from a theoretical perspective: for example, theories are often built on \emph{coarse-grained} descriptions of high-dimensional biological systems~\cite{durrett1994tpb}, an approach inspired by physics and applied mathematics. These latent dimensions can be found either based on such coarse-grained theories about the system and/or biologically meaningful quantities: in the context of fish schooling, inspired by mesoscopic theories of collective motion~\cite{biancalani2014prl, jhawar2019bookchapter}, we used the 2-dimensional polarization vector as the coarse-grained variable. Alternatively, one can use data-driven techniques to identify latent variables~\cite {greenacre2022pca,schmid2022dmd,van2009dimensionality}. The problem of jointly discovering a suitable latent subspace and dynamical equations in that subspace has been explored for deterministic dynamics~\cite{champion2019data}; developing similar approaches for stochastic systems is a worthwhile direction for future work.

However, there may be situations when the system dynamics require a high-dimensional description or when the observed dimensions are insufficient to fully capture the dynamics. Here, an SDE model---or any low-dimensional model---in terms of only the observed variables cannot capture the full dynamics of the system. This naturally leads to the question of how do we evaluate potentially erroneous applications of our easy-to-use tool. In this context, the diagnostic tools of PyDaDDy are crucial to make an informed judgment about the discovered model. Specifically, when there are unobserved dynamical variables, the residual autocorrelation may have a longer decay time-scale, suggesting that there are residual slow dynamics in the system that are unexplained by the SDE model. Similarly, the model diagnostics may fail to produce simulated histograms or autocorrelation functions that match the originals. This was observed when we tried to model the cell dynamics using an overdamped SDE, while the actual dynamics could not be described using the position dimension alone (SI Section S4 B). Another example is illustrated in section SI Section S5 B, where we attempt to describe the dynamics of a two-species interaction model from the time series of a single species, where the diagnostic tests revealed that the reduced dimensional model is incomplete.


\paragraph{Estimation with limited data} As is true for any estimation method, data-driven SDE discovery also requires sufficiently good quality data. Our analysis showed that both the short length of the data and low sampling frequency lead to inaccurate SDE estimations. The silver lining, however, was that the estimated model did not yield good diagnostics, which helps the user to be cautious about further interpretations of the model. Generally, the amount of data required for a satisfactory reconstruction is governed by the complexity of the underlying model and the level of measurement noise~\cite{fajardo2023fundamental}. While providing exact numbers for the amount of data is difficult, we do a numerical exploration of the estimation performance of PyDaDDy with limited data in SI Section S5 A. The highlight of this analysis is that equation discovery techniques will outperform conventional bin-wise Kramers-Moyal averaging, especially when the available data is limited. 

In the presence of measurement noise, we expect the approach to work well in the regime where the measurement noise is sufficiently small. In this regime, we expect that the estimate of the drift function will be unaffected on average. In contrast, the diffusion estimates are likely to be biased by an amount equal to the measurement noise variance.~\cite{bottcher2006reconstruction}; this may be helpful when the focus is on building predictive models. On the other hand, if the focus is on developing mechanistic models, one would need to separate measurement noise explicitly from the underlying biological processes. More broadly, how different types of noise structures -- correlated across components, correlated over time and the measurement noise -- affect the SDE inference offer interesting avenues for further research. Finally, the case of sporadic missing data points in the time series (e.g. due to sensor failures) is less serious. The SDE estimation can proceed with only the available data points without issues, as shown in the fish dataset. 



\subsection*{Related work and Concluding remarks}

Equation learning was first introduced in the engineering literature in the context of ordinary and partial differential equations and is implemented in popular packages like \emph{PySINDy}~\cite{desilva2020pysindy} and~\emph{DataDrivenDiffEq.jl}~\cite{datadrivendiffeq}. These methods, however, are restricted to discovering \emph{deterministic} differential equations. There are some exceptions, such as the R package \emph{Langevin}~\cite{rinn2016langevin}, which takes a (one or two-dimensional) time series as input and estimates the drift and diffusion coefficients as bin-wise averaged Kramer-Moyal coefficients. Some functionality is provided for visualizing and diagnosing the estimated coefficients. However, these protocols do not produce interpretable SDEs; furthermore, bin-wise averaged estimates can often be error-prone~\cite{jhawar2020inferring,callaham2021langevin} especially when dealing with limited data (SI Section S5 A)---this issue is largely mitigated in our equation learning approach. 

We emphasize that most tools of data-driven models, including in the context of dynamical systems  models~\emph{DiffEqParamEstim.jl}~\cite{DifferentialEquations.jl-2017} and \emph{Sim.DiffProc}~\cite{guidom2020diffproc}, take the classic approach where the user specifies a parametric form of the drift and diffusion functions. In such cases, the user must begin with a set of candidate models and then use selection procedures to find the suitable dynamical system model. Typically, the model selection criterion for stochastic dynamical models is based on how well the steady state distribution of the state variable matches with the real data. Such methods, however, cannot \textit{discover} the fact that the time series in Fig.~\ref{fig:synthetic}A-\textit{i} and Fig.~\ref{fig:synthetic}A-\textit{ii} that have nearly identical distributions but come from fundamentally differing dynamical models. In contrast, our approach can not only discover SDEs directly from time series data but can also correctly distinguish between contrasting governing equations. 



We are making rapid headway towards collecting large datasets in biological systems across scales. As the era of big data looms over biology, we hope that our presentation of a method together with a relatively easy-to-use package -- that helps one characterize the governing dynamical equations from the data -- will inspire the usage of these methods in the field. We can thus exploit the potential for mechanistic predictive models while accounting for the inherent stochasticity of the systems. We expect many challenges, and hence opportunities for future work, when applying these methods to other complex real-world datasets that may be high-dimensional, driven by changing parameters, influenced by observational noise and complicated noise structures (including coloured and non-Gaussian noise).

\section*{Code and data availability}

The code for data-driven SDE discovery is available as a Python package at \url{https://github.com/tee-lab/PyDaddy} (archived at \url{https://doi.org/10.5281/zenodo.13777396}~\cite{pydaddy-zenodo}), along with detailed documentation and tutorials at \url{https://pydaddy.readthedocs.io.} Several tutorial notebooks are provided at \url{https://pydaddy.readthedocs.io/en/latest/tutorials.html} to familiarize the user with the package.

The fish schooling dataset is from~\cite{jhawar2020fish}, and is available at \url{https://doi.org/10.5281/zenodo.3596324}~\cite{jhawar2020fishdataset}. A subset of this dataset is provided as sample data with the package.

The cell migration dataset is from~\cite{bruckner2019stochastic}. The dataset is provided as sample data with the package.

\section*{Acknowledgements}
VG acknowledges support from the Science and Engineering Research Board, Department of Biotechnology, and Indo-French Centre for the Promotion of Advanced
1036 Research (64T4-1). DRM acknowledges support from the DST INSPIRE Faculty Award. JJ acknowledges support from the Humboldt Postdoctoral Fellowship and the Heidelberger Akademie der Wissenschaften, Heidelberg, Germany. DBB acknowledges support from the NOMIS foundation and the EMBO Postdoctoral fellowship (ALTF 343-2022). AN and SP acknowledge support from the MoE PhD Fellowship. The authors thank Ashrit Mangalwedhekar, Vivek Jadhav, Shikhara Bhat, Cassandre Aimon and Harishankar Muppirala for comments on the manuscript and code. We thank Kollegala Sharma for his input on the Kannada translation of the title and abstract. 

\section*{Author contributions}
\paragraph*{Arshed Nabeel} Data curation, Formal analysis, Investigation, Methodology, Software, Visualization, Writing (Original Draft Preparation), Writing (Review and Editing)

\paragraph*{Ashwin Karichannavar} Data curation, Formal analysis, Investigation, Methodology, Software

\paragraph*{Shuaib Palathingal} Investigation, Validation

\paragraph*{Jitesh Jhawar} Data curation, Investigation

\paragraph*{David Br\"{u}ckner} Data curation, Investigation

\paragraph*{Danny Raj M} Conceptualization, Project Administration, Validation, Writing (Original Draft Preparation), Writing (Review and Editing)

\paragraph*{Vishwesha Guttal} Conceptualization, Funding acquisition, Project Administration, Resources, Supervision, Validation, Writing (Original Draft Preparation), Writing (Review and Editing)

\bibliography{references}
\bibliographystyle{amnatstyle.bst}

\end{document}



\title{Supplementary Information: Discovering stochastic dynamical equations from ecological time series data}

\author{Arshed Nabeel}
 \email{arshed@iisc.ac.in}
\affiliation{IISc Mathematics Initiative, Indian Institute of Science, Bengaluru, India}
\affiliation{Center for Ecological Sciences, \\
             Indian Institute of Science, Bengaluru, Karnataka, 560012, India}

\author{Ashwin Karichannavar}
\affiliation{Center for Ecological Sciences, \\
             Indian Institute of Science, Bengaluru, Karnataka, 560012, India}

\author{Shuaib Palathingal}
\affiliation{Center for Ecological Sciences, \\
             Indian Institute of Science, Bengaluru, Karnataka, 560012, India}
             
\author{Jitesh Jhawar}
 \altaffiliation[Now at: ]{School of Arts and Sciences, Ahmedabad University, Ahmedabad, India.}
\affiliation{University of Konstanz, Konstanz, Germany}
\affiliation{Max Planck Institute of Animal Behaviour, Konstanz, Germany}

\author{David B. Brückner}
\affiliation{Institute of Science and Technology, Austria}
\affiliation{Biozentrum, University of Basel, Switzerland}

\author{Danny Raj M.}
   \email{danny@iitm.ac.in}
\affiliation{Department of Chemical Engineering, \\
             Indian Institute of Science, Karnataka, 560012, India}
\affiliation{Dept of Applied Mechanics and Biomedical Engineering, \\ 
IIT Madras, Chennai, 600036, India.}

\author{Vishwesha Guttal}
 \email{guttal@iisc.ac.in}
\affiliation{Center for Ecological Sciences, \\
            Indian Institute of Science, Bengaluru, Karnataka, 560012,India}






\maketitle
\renewcommand{\thesection}{S\arabic{section}}
\renewcommand{\thefigure}{S\arabic{figure}}
\renewcommand{\thetable}{S\arabic{table}}











\tableofcontents

\newpage
\setlength{\tabcolsep}{12pt}
\begin{table*}[]
    \centering
    \begin{tabular}{cl}
        \toprule
         $\bx(t)$   & A $d$-dimensional time series of a state-variable. \\
         $\bff(\bx)$  & Deterministic component of the SDE (drift function). \\
         $\bg(\bx)$   & Stochastic component of the SDE (diffusion function). \\
         $\bG(t)$   & $\bg \bg^T$, also sometimes called the diffusion function. \\
         $\boldeta(t)$  & $d$-dimensional Gaussian white noise process. \\
         $F(t; \bx)$    & Instantaneous estimate of the drift at time $t$ \\
         $G(t; \bx)$    & Instantaneous estimate of the diffusion at time $t$ \\
         $\boldeps(t)$ & Estimated residual at time $t$, based on which $G$ is computed. \\
         $f$            & Any component of $\bff$ or $\bG$. \\
         $f_i$          & Candidate terms in the library for sparse regression. \\
                        & $f$ is estimated as a sparse linear combination of $f_i$'s. \\
         $\lambda$      & Sparsification threshold, a hyperparameter in the equation learning algorithm. \\
         $\hat{\bff}, \hat{\bG}$ & Estimates of the drift and diffusion functions $\bff$ and $\bG$ respectively. \\
         $\brr(t)$      & Estimate of $\boldeta(t)$ at time $t$ based on the estimated SDE. \\
         \midrule
         $\bm$          & The 2-D polarization vector for the fish schooling dataset. \\
         $x, v$         & The position and velocity respectively for the cell migration dataset. \\
         \midrule
         $d$            & Dimension of the state variable. \\
         $T$            & Number of time points in the observed dataset. \\
         $X$            & A $T \times d$ matrix containing observed data points. \\
         $\boldsymbol{\phi}$ & A $T \times d$ matrix containing $f(\bx)$ evaluated at each time point, for some component $f$. \\
         $\Theta$       & A $T \times k$ matrix containing some component of the instantaneous drift or diffusion. \\
         $\boldsymbol{\xi}$ & Unknown variable to be estimated in the sparse regression problem, $\theta $ \\
         \bottomrule
    \end{tabular}
    \caption{Table of key symbols and notation.}
    \label{tab:g_coeffs}
\end{table*}

\newpage

\section{Technical details of the estimation procedure}

\subsection{Sparse regression with sequentially thresholded least-squares (STLSQ)}
Recall that we had an observed time series $\bx(t)$ and the instananeous estimates of drift and diffusion functions $F(t; \bx)$ and $G(t; \bx)$. We need to estimate expressions for the drift and diffusion functions component-wise.

Let $X$ be the $T \times d$ matrix containing the observed time series data, where each row is one time-point. To estimate the expression for a component $f$ of the drift or diffusion, $\boldsymbol \phi$ be the $T \times 1$ vector containing the corresponding components of the instantaneous drift or diffusion function. The algorithm proceeds as follows:
\begin{enumerate}
\item Construct the $T \times k$ matrix $\Theta = [ f_1(X) \, f_2(X) \, \cdots \, f_k(X) ]$ where $f_i(X)$ is a shorthand for applying $f_i$ row-wise on $X$.
\item Solve the least-squares regression problem, $\boldsymbol \phi = \Theta \boldsymbol \xi$ for the $k$-dimensional unknown vector $\boldsymbol \xi$.
\item For an appropriately chosen (see below) \emph{sparsification threshold} $\lambda$, set all $\boldsymbol \xi_i < \lambda$ to $0$, and drop the corresponding columns from $\Theta$.
\item Repeat the above procedure iteratively, until no more columns can be eliminated.
\end{enumerate}

\subsection{Advanced diagnostic techniques}

The \emph{Methods} section (subsection \emph{Diagnostics}) of the main text discussed several diagnostic tests that can be used to validate the discovered SDE model. More advanced diagnostic tests are also available, which test for further assumptions involved in modelling a stochastic process using an SDE. Some of these are implemented in the PyDaddy package, and are described below:

\begin{itemize}
    \item Recall that the residuals $\boldeps(t)$ are an estimate for the noise $\boldeta(t)$, and should have a $d$-dimensional standard normal distribution. The deviation of the distribution of $r(t)$ from the standard normal distribution can be examined using QQ-plots for the individual components of $\boldeps$.
    \item \emph{Pawula's Theorem:}  Pawula's theorem states that the third and higher-order Kramers-Moyal (KM) coefficients 
    should be zero, if a process can be modelled using an SDE~\cite{gardiner2009}. In practical scenarios where we have finite-length time series sampled at a finite sampling time, the alternative to Pawula's theorem is that $K_4(x) \approx 3 \cdot K_2(x)^2$, where $K_2$ and $K_4$ are the second and fourth KM coefficients
    respectively~\cite{tabar2019book, lehnertz2018characterizing}. This is applicable only for a one-dimensional time series and not for higher dimensions.
\end{itemize}

SDEs discovered from real-world datasets may not always satisfy all these diagnostic tests. For descriptive models, it is often satisfactory if the model passes the simple noise and model diagnostic tests mentioned in the \emph{Methods} section (subsection \emph{Diagnostics}). However, for quantitative applications where predictive accuracy is important, one has to ensure that all the assumptions behind SDE models are met---hence, the advanced diagnostic tests mentioned here become crucial. 

\section{PyDaDDy Package: a brief introduction}
\label{ax:pydaddy}

We implement our technique for data-driven SDE discovery as a Python package, \textbf{PyDaDDy} ({\bf Py}thon library for {\bf Da}ta {\bf D}riven {\bf Dy}namics). PyDaDDy takes either a scalar or 2-dimensional vector time series as input. Although the theory of jump moment estimation and equation learning applies to arbitrary dimensions, visualising and diagnosing results becomes infeasible for dimensions larger than two. Therefore, we choose to limit the functionality of the package to dimensions up to two.


From a given uniformly sampled 1-D or 2-D time series, PyDaDDy can compute, visualise, and fit drift and diffusion functions. PyDaDDy also supports time series with missing data points. Diagnostic tools are provided to verify whether the assumptions necessary for modelling the time series as an SDE are met. Finally, PyDaDDy can export data into a DataFrame or a CSV file. Fig.~1 of main text shows an overview of the package and its various functionalities.


For ease of use, we make the package and the associated tutorial notebooks accessible online, even without installation, via Google Colab: \url{https://pydaddy.readthedocs.io/en/latest/usage.html#pydaddy-on-google-colab}. 

The package can also be installed on a local machine using PIP or Conda:

\begin{minted}{bash}
pip install pydaddy
\end{minted}

or

\begin{minted}{bash}
conda install pydaddy -c tee-lab
\end{minted}

PyDaDDy features a single-command operation mode, which allows the user to run a dataset through PyDaDDy and generate a comprehensive HTML report. This can be done using the command: 

\begin{minted}{bash}
pydaddy <dataset-file-name>
\end{minted}

The report contains interactive figures of the estimated drift and diffusion functions (with default parameters and automatic threshold tuning---see below) and diagnostic results about noise statistics and model self-consistency. 

PyDaDDy also has a comprehensive Python API for advanced usage; which allows the user to individually generate various plots, perform diagnostics, or fine-tune the function fitting procedure.

\begin{figure*}
    \centering
    \includegraphics[width=\textwidth]{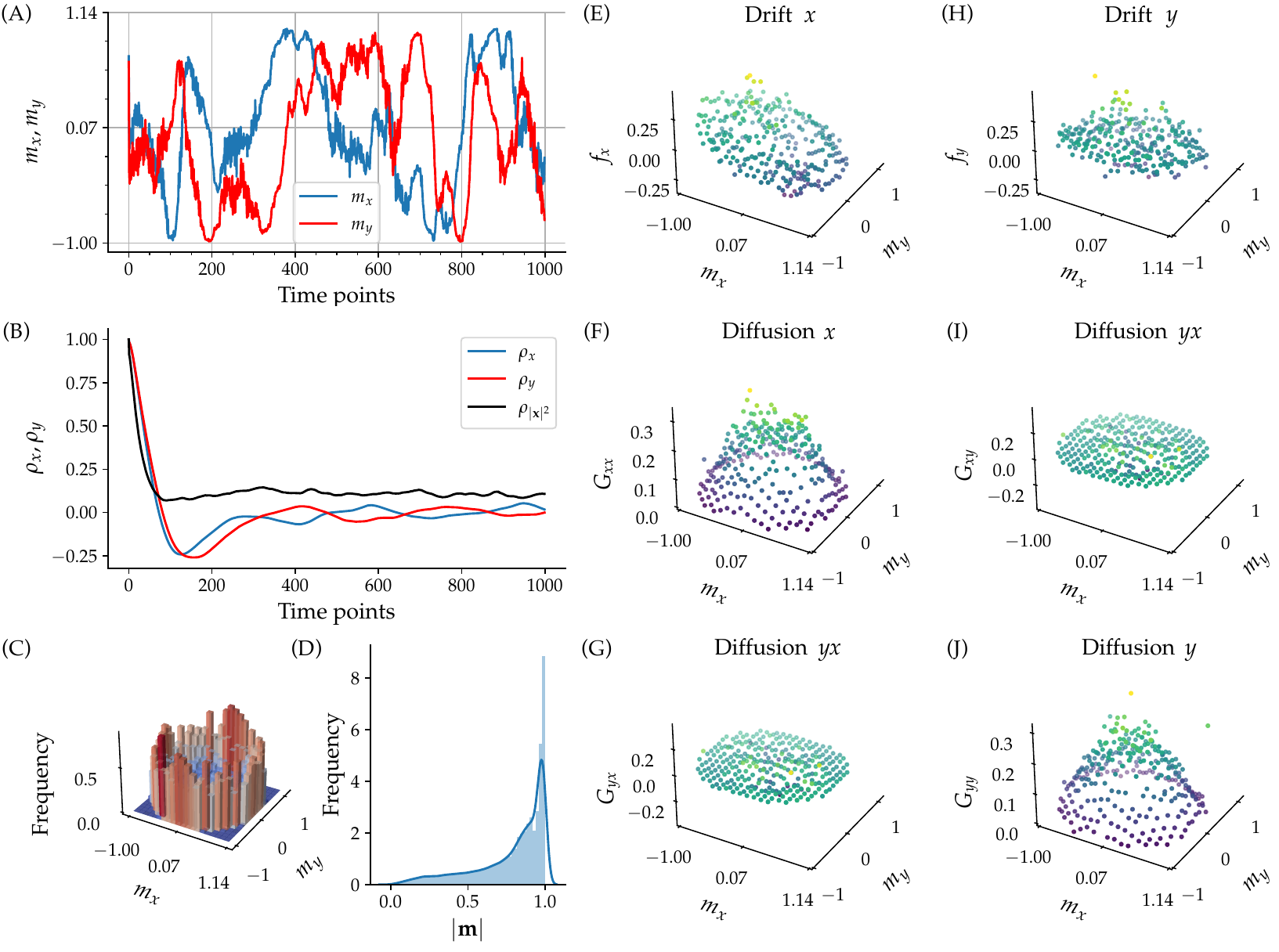}
    \caption{{\bf The summary plot generated by PyDaddy when applied to schooling fish data from the reference~\cite{jhawar2020fish}}.  
    (A) Input time series (fish group polarization), a vector time series denoted by $\mathbf m = (M_x,M_y)$. One time step corresponds to 0.12 s. (B) Autocorrelation functions of the time series, showing that the autocorrelation decays with time. (C, D) Frequency histograms of time series $\mathbf m$ and $| \mathbf m |$.
    (E, H) Preliminary estimates of drift functions. (F, G, I, J) and diffusion functions. The drift and diffusion functions shown here are bin-wise averages of the jump moments (bin-wise averaging is done only for the visualisation).
    }
    \label{fig:estimation-summary}
\end{figure*}

\subsection*{Summary figure and estimation of drift and diffusion}

The estimation workflow starts by creating a PyDaDDy object \texttt{dd}, initialised with a dataset.

\begin{minted}{python}
dd = pydaddy.Characterize([x, y], t=0.12)
\end{minted}

where \texttt{x} and \texttt{y} are Numpy arrays of the $m_x$ and $m_y$ time series respectively. A summary figure is generated with basic statistics of the time series such as the autocorrelation function and histograms of  $\mathbf m$. The summary figure also shows a preliminary estimate of the drift and diffusion functions, using binwise averaged jump moments --- see~\cite{jhawar2020inferring} for more details on the bin-wise averaging method. 


\begin{figure}
    \centering
    \includegraphics[width=\linewidth]{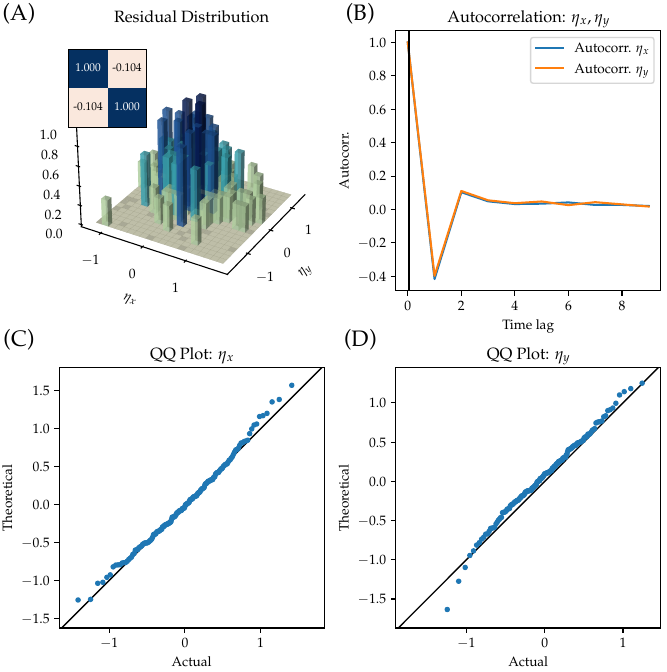}
    \caption{{\bf Noise diagnostics figure generated by PyDaDDy when applied to schooling fish data from the reference~\cite{jhawar2020fish}} (A) Distribution of the residuals, which resembles a bivariate Gaussian with zero mean. The inset matrix shows correlation matrix of the $x$ and $y$ components of the noise. This should be close to the identity matrix. (B) Autocorrelation functions of the residuals, with the black line marking the autocorrelation time and is close to zero, as desired. (C, D) QQ-plots of the marginal distributions, which nearly falls on a line of slope 1, again as desired by the assumptions of the SDE model.
    }
    \label{fig:noise-diag}
\end{figure}

\begin{figure*}
    \centering
    \includegraphics[width=\textwidth]{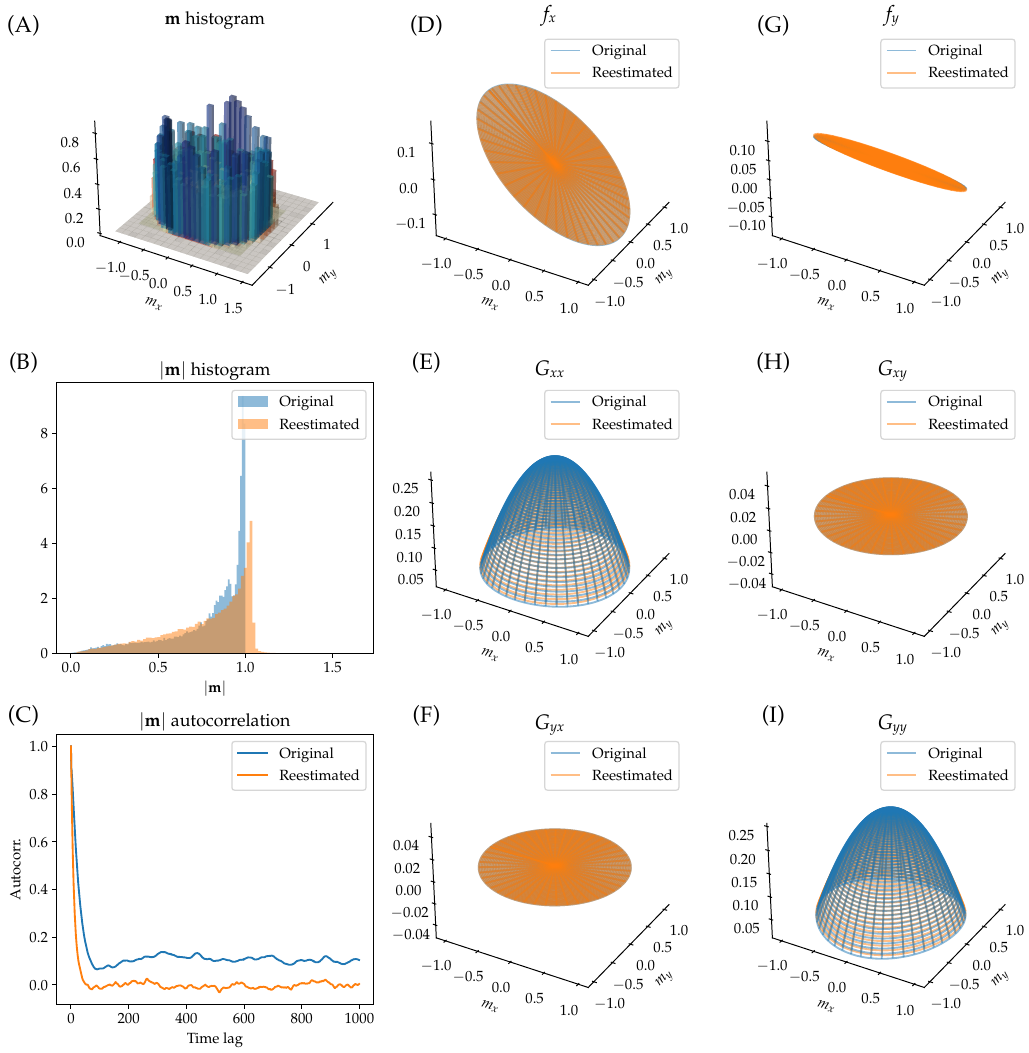}
    \caption{\textbf{Model diagnostics figure generated by PyDaddy when applied to schooling fish data from the reference~\cite{jhawar2020fish}}
    (A-B) Comparison of histograms of $M$ and $|M|$ respectively, between the original time series and a time series simulated using the discovered SDE. (C) Autocorrelation functions of the simulated time series. (D-I) Comparison of the drift and diffusion functions between the original estimate and re-estimate (from the simulated data).}
    \label{fig:model-diag}
\end{figure*}

\subsection*{Fitting analytical expressions for drift and diffusion functions}

PyDaDDy uses a sparse regression technique for model fitting for the drift and diffusion functions. This can be done using the \texttt{fit()} function of the \texttt{dd} object. By default, PyDaDDy fits polynomial functions, but other function families can also be fit using custom libraries. A \emph{sparsification threshold}, which governs the number of terms in the fit (see SI Section S1B) can be either chosen automatically using cross-validation (see SI Section S1C), or be set manually.

For example,

\begin{minted}{python}
dd.fit('F1', order=3, tune=True, plot=True)
\end{minted}

\noindent fits a degree-3 polynomial (\texttt{order=3}) with sparsification threshold chosen automatically (\texttt{tune=True}). The first argument names the function to be fitted: \texttt{F1} and \texttt{F2} correspond to the two components of the drift; \texttt{G11} and \texttt{G22} correspond to diffusion, and \texttt{G12} and \texttt{G21} correspond to cross diffusion (Notice the slight difference in notation, where the package uses uppercase $F$ and $G$ to denote the drift and diffusion functions). The \texttt{plot=True} argument tells the function to produce a cross-validation error plot for a range of thresholds. The ideal threshold should strike a balance between the error and number of terms in this plot. PyDaDDy chooses as high as a threshold as possible without increasing the cross-validation error too much.


The threshold can also be manually set by the user. For example,

\begin{minted}{python}
dd.fit('F1', order=3, threshold=0.1)
\end{minted}

produces a polynomial with degree up to 3, with only terms with coefficients 0.1 or higher.

\subsection*{Diagnostics}
It is essential to ensure that the assumptions involved in modelling the time series as an SDE are not violated. PyDaDDy provides diagnostic functions to verify that the assumptions involved in modelling a time series as an SDE are met (see \emph{Methods} in the main text). Diagnostic plots in PyDaDDy can be generated using
\begin{minted}{python}
dd.noise_diagnostics()
\end{minted}

SI~Fig.~\ref{fig:noise-diag} shows the output of the noise diagnostics for the fish schooling dataset.

In addition, PyDaDDy can also perform a model self-consistency check (see \emph{Methods}) using the following command:
\begin{minted}{python}
dd.model_diagnostics()
\end{minted}

This function uses the estimated SDE to generate a simulated time series, with the same length and sampling interval as the original dataset. It then estimates the drift and diffusion functions from the simulated time series, using the same order and threshold used in the original estimation. Finally, a summary figure is generated, comparing the histogram, drift and diffusion functions of the original dataset to the simulated dataset (SI~Fig.~\ref{fig:model-diag}). If the model is self-consistent, we expect a good match between the original and the simulated time series.


\section{Demonstration with classic models in ecology}
\label{ax:models}

In this section, we demonstrate the generality of the our protocol (via the PyDaddy package) with several classical models in theoretical ecology and beyond. The models considered are listed below.

\paragraph{Logistic model of population growth.} We consider the logistic model of population growth, with a demographic noise term that grows as the square root of the population size, governed by the following SDE:

\begin{align}
    \dot N = 2N \left( 1 - \frac{N}{5} \right) + \sqrt{N} \cdot \eta(t)
\end{align}

\paragraph{Harvesting model.} We consider a population growth model with a harvesting term of Holling's Type III form, with a multiplicative noise term.
\begin{align}
    \dot N = 2.4 N \left( 1 - \frac{N}{6} \right) - \frac{4N^2}{1 + N^2} + 0.2 \left( \frac{N^2}{1 + N^2} \right) \cdot \eta(t)
\end{align}

\paragraph{Lake eutrophication model.} We consider a lake eutrophication model ~\cite{carpenter1999management} with an additive noise term.
\begin{align}
    \dot x = 0.5 - x + \frac{x^8}{1 + x^8} + 0.2 \cdot \eta(t)
\end{align}

\paragraph{Lotka-Volterra competition model.} We consider the classical Lotka-Volterra model for inter-species competition.

\begin{align}
    \dot x &= 2x \left( 1 - \frac16 (x + y) \right) + 0.1 x \cdot \eta(t) \\
    \dot y &= 4y \left( 1 - \frac18 (2x + y) \right) + 0.1 y \cdot \eta(t) \label{eq:lv-comp}
\end{align}

\paragraph{Van der Pol oscillator.} We considered the Van der Pol model of relaxed, non-linear oscillations, with the velocity dynamics being perturbed by an additive noise term.

\begin{align}
    \dot x &= v \\
    \dot v &= -x + 5(v + x^2v) + \eta(t)
\end{align}

\paragraph{Prey-predator model.} Finally, we considered a Prey-predator model ~\cite{alonso2002mutual} with predation having a Holling's Type II form.

\begin{align}
    \dot n &= n(n-1)-\frac{np}{n+p} + 0.1 n \cdot \eta(t) \\
    \dot p &= 0.48\frac{np}{n+p}-0.2p + 0.1 p \cdot \eta(t)
\end{align}

For each model, we generated simulated time series data (for $10^5$ time points with $\Delta t = 0.01$) using the corresponding SDE model using an Euler-Maruyama SDE integration scheme. We then used the simulated time series to estimate SDEs using PyDaDDy. The results of PyDaDDy SDE discovery are given in Table 1 of main text. 

For the logistic, Lotka-Volterra and Van der Pol models, we used the default polynomial library for fitting, as these are indeed polynomial models. For the other models, we used custom libraries. The library consisted of terms $x, x^2, x/(x+1), x^2/(x^2+1)$ for the harvesting model, $1, x, x^2, x^8/(x^8+1)$ for the lake eutrophication model, $n, p, n^2, p^2, np/(n+p)$ for the prey-predator model. For the harvesting and Lotka-Volterra simulations, we generated 10 different simulations with different initial conditions, so that we get a good coverage of the state-space (see SI Section S5for more on this point).

For all the models, see that PyDaDDy is able to recover SDEs that closely match the original ground-truth models (see \emph{Table 1} in the main text).

We pause here to make a few general notes on constructing a custom library. First, the choice of basis functions in the library should be informed by some knowledge of the underlying physics of the problem. A second and equally important point is that the functions in library should be reasonably orthogonal with respect to each other (i.e., for two functions $f_1(x)$ and $f_2(x)$, the \emph{inner product} $\int f_1(x)f_2(x) dx$ should be close $0$ over the relevant ranges). If this is not the case, it becomes difficult for regression to disentangle individual contributions of the different functions in the library. For example, for the lake eutrophication model, if the library contains terms $x^a/(1+x^a)$ for many different values of $a$, all these functions are close to each other and sparse regression may end up representing the harvesting term as a combination of these. A better approach (which we followed in our analysis) would be to try many different fits, each with a library containing only one functional response term, and choose the model with the fit that minimizes the overall error.

\section{Model selection with real-world datasets}
\label{ax:model-selection}

Here, we discuss a typical workflow for discovering data-driven equations from real-world datasets. The process typically involves going back and forth between model discovery and model diagnostics to get the simplest, \emph{`good enough'} model that captures the essential features of the dynamics. For simplicity, we discuss the procedure in the context of discovering polynomial models. 

The model discovery procedure has two \emph{hyperparameters}: the degree of the polynomial, which governs the total number of terms in the library, and the sparsification threshold, which governs how many terms from the library actually end up in the discovered function. Loosely, the model discovery procedure will then have the following steps:

\begin{enumerate}
    \item Based on visual inspection, the binned drift and diffusion coefficients, choose an appropriate polynomial degree. While there are limitations in using the binned estimates (see \emph{Methods}, section \emph{Contrast to conventional approaches}), they serve as a starting point to visualize the structure of the drift and diffusion functions. The fitting procedure does not depend on the binned estimates.
    \item Set the sparsification threshold to be 0 (i.e., a polynomial of the specified order with no sparsification will be returned).
    \item Generate simulated time series with the discovered models and compare the statistics (histograms and autocorrelation functions) of the simulated time series with the original time series.
    \item If the histogram or autocorrelation match sufficiently well, we can conclude that the polynomial degree is high enough to capture the model dynamics. Now, progressively increase the threshold and find the maximum value of threshold (i.e. the sparsest model of the specified degree) that captures the dynamics sufficiently well.
    \item If the simulated histogram or autocorrelation do not match the original data, go back to step 1, increase the polynomial degree, and repeat the procedure.
    \item Finally, ensure that the chosen model is self consistent. That is, starting with a simulated time series generated with the model, the model discovery procedure should produce the same model.
\end{enumerate}

It is important to note that the cross validation approach to model selection (\emph{Methods}, section \emph{Model selection for sparse regression}) also provides a means to identify a sparse model. However, this is only based on how well the inferred model explains the data that is held back for testing. The manual approach detailed above checks for model consistency, and if the model is able to reproduce the histograms and autocorrelation of the state variable, which weren't used in the identification of the SDE model.

\subsection{Model selection for the fish schooling dataset}

\begin{figure}
    \centering
    \includegraphics[width=0.6\linewidth]{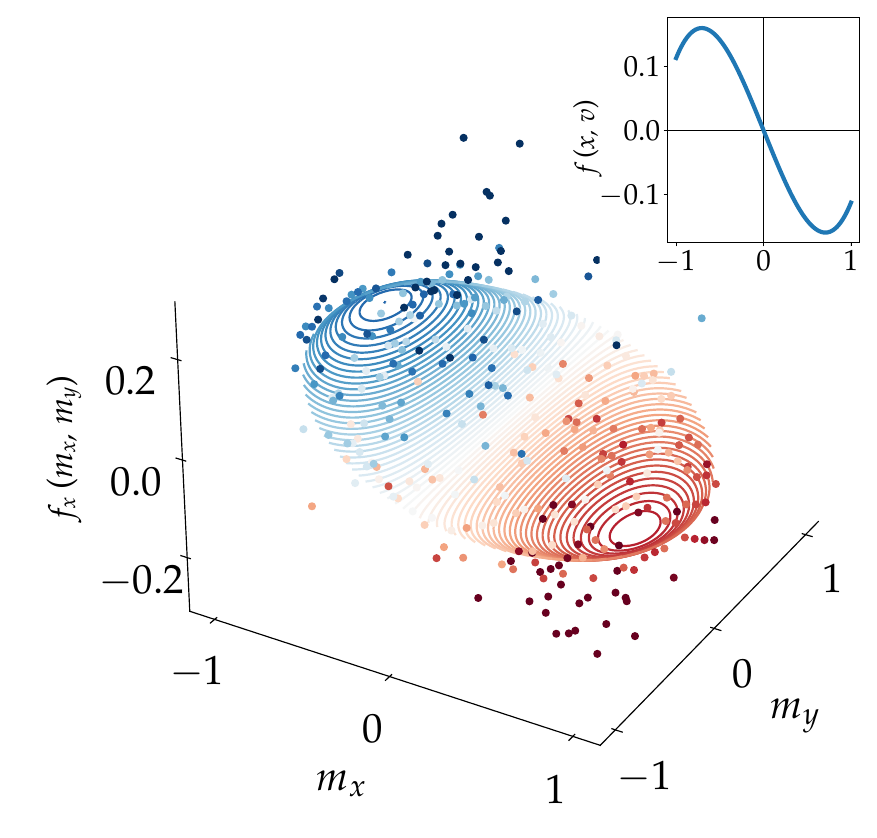} 
    \caption{A cubic function offers a slightly better fit for the drift.
    }
    \label{fig:cubic-drift}
\end{figure}

\begin{figure*}
    \centering
    \includegraphics[width=\textwidth]{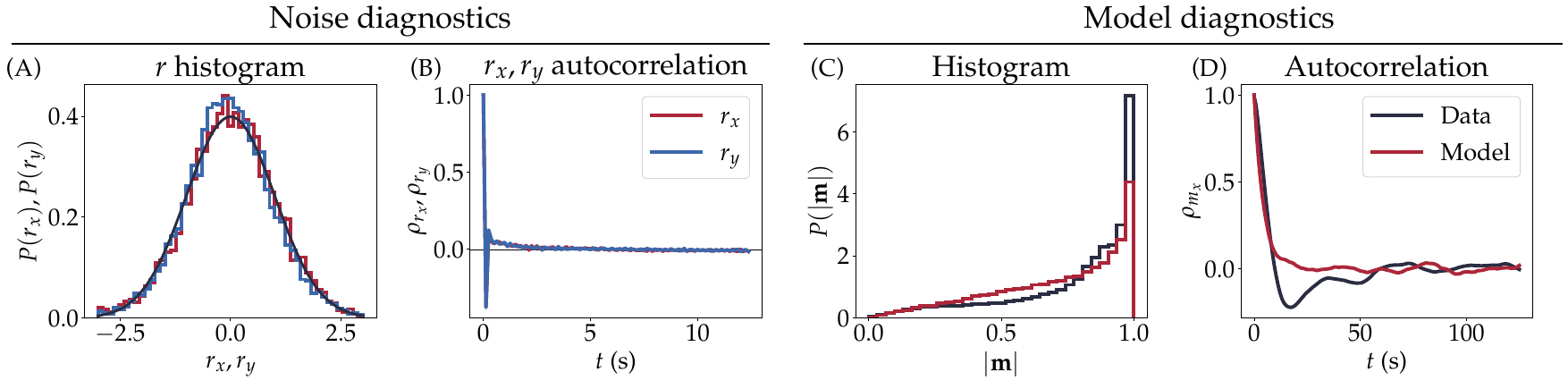} 
    \caption{Noise and model diagnostics, with an SDE model with a cubic drift function. This model offers little to no improvement over an SDE with linear drift.
    }
    \label{fig:cubic-drift-diag}
\end{figure*}

For the fish schooling dataset, $\bm(t)$ is a vector, so the governing SDE will be a vector SDE. Therefore, the drift $\bf$ is a vector function and $\bg$ is a matrix function. Sparse fitting can be performed on their individual components independently. Note that it is possible to take in to account the inherent symmetries of $\bm$ to constrain the possible terms and directly derive a vector SDE---this procedure is described in more detail in~\cite{nabeel2023data}.

Based on visual examination, the components of the drift function $f_x$ and $f_y$ cross 0 at $\bm = 0$, the most parsimonious function that captures this being a linear function. On the other hand, the diffusion components $g_{xx}$ and $g_{yy}$ have a dome shape, captured by a quadratic. The cross-terms in the diffusion function, $g_{xy}$ are negligible and can be safely ignored.

Model diagnostics (main text, Fig.3) reveals that the histogram of the simulated time series matches closely with that of the original time series. The autocorrelation matches the decay envelope of the autocorrelation but not the oscillations. The oscillations in the autocorrelation function are caused by the school swimming along the boundary of the tank and is a superfluous effect we are not interested in capturing in our minimal SDE model. Therefore, we conclude that a model with a linear drift function, quadratic diffusion function and no cross-diffusion terms is sufficient to capture the essential dynamics of the polarization dynamics of karimeen fish.

It's worth noting here that a cubic function can give a slightly better fit for the drift function (Fig.~\ref{fig:cubic-drift}). However, the cubic function does not change the stability landscape by adding new equilibrium manifold. Furthermore, as shown by the diagnostics (Fig.~\ref{fig:cubic-drift}), this model performs no better in terms of replicating the essential statistical features of the data. Therefore, keeping parsimony and model interpretability in mind, we choose the linear drift function over the cubic function.

\subsection{Model selection for the cell migration dataset}
\subsubsection*{An overdamped model is insufficient}
\begin{figure*}
    \centering
    \includegraphics[width=\textwidth]{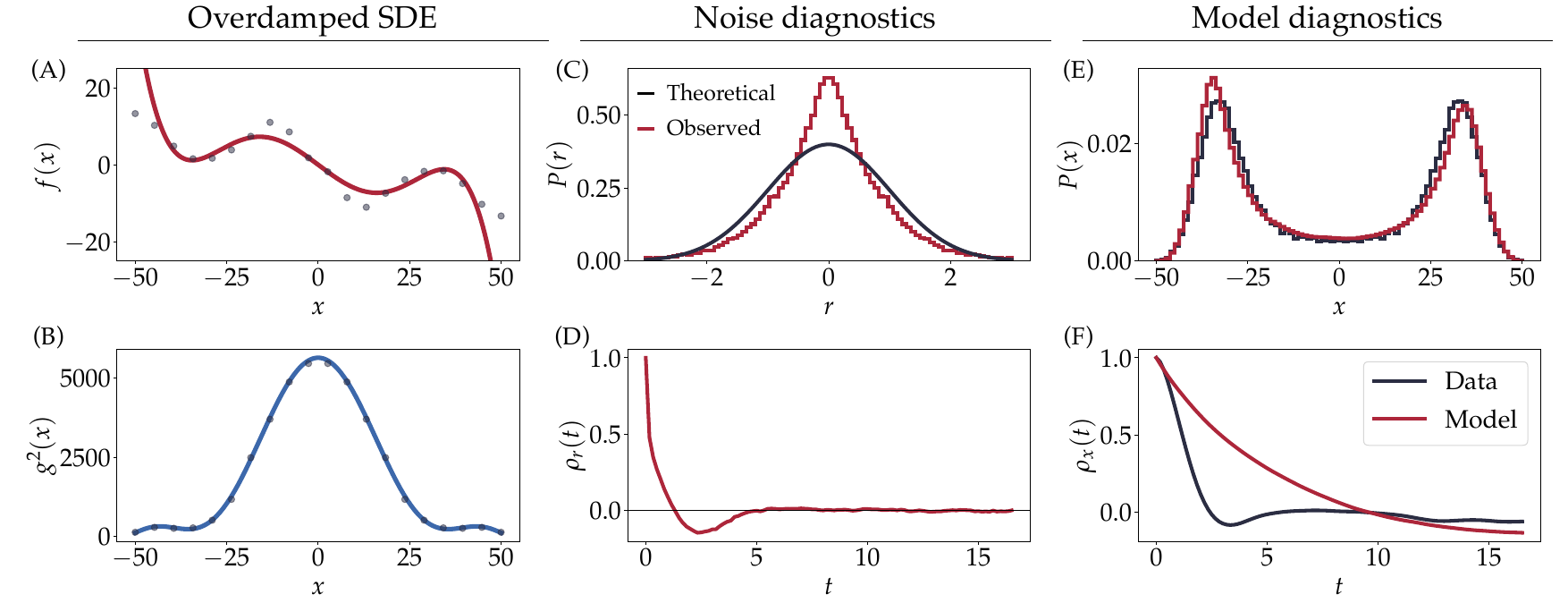} 
    \caption{An overdamped model is insufficient to capture the cell dynamics. Note subplots (D), (F) which show that the estimated model is insufficient. 
    (A, B) The discovered drift and diffusion functions. Bin-wise averaged estimates are also shown. (C) Histogram of the residuals (red) and the theoretically expected standard normal distribution (black). (D) Autocorrelation of the residual, showing a long decay time. (E, F) Histogram and autocorrelation of model time series, plotted alongside the observed data.
    }
    \label{fig:cell-mig-overdamped}
\end{figure*}

We first attempt to model the cell trajectories using an overdamped SDE of the form
\begin{align}
    \dot x = f(x) + g(x) \cdot \eta(t) \label{eq:cell-eqn-overdamped}
\end{align}

The drift and diffusion functions are fitted using a fifth and eighth order polynomial respectively (Fig.~\ref{fig:cell-mig-overdamped}). However, diagnostic tests make it evident that the underdamped model is insufficient to fully explain the observed dynamics. Specifically, there is a mismatch between the autocorrelation functions of the original time series and model simulations (panel D). In addition, The autocorrelation of the estimated residuals $r(t)$ has a long decay time, violating the white-noise assumption. This suggests that there are dynamics in the system that the current model, which is in terms of only the position as a state variable, is unable to capture.

\subsubsection*{An underdamped model for cell dynamics}

\begin{figure}
    \centering
    \includegraphics[width=\linewidth]{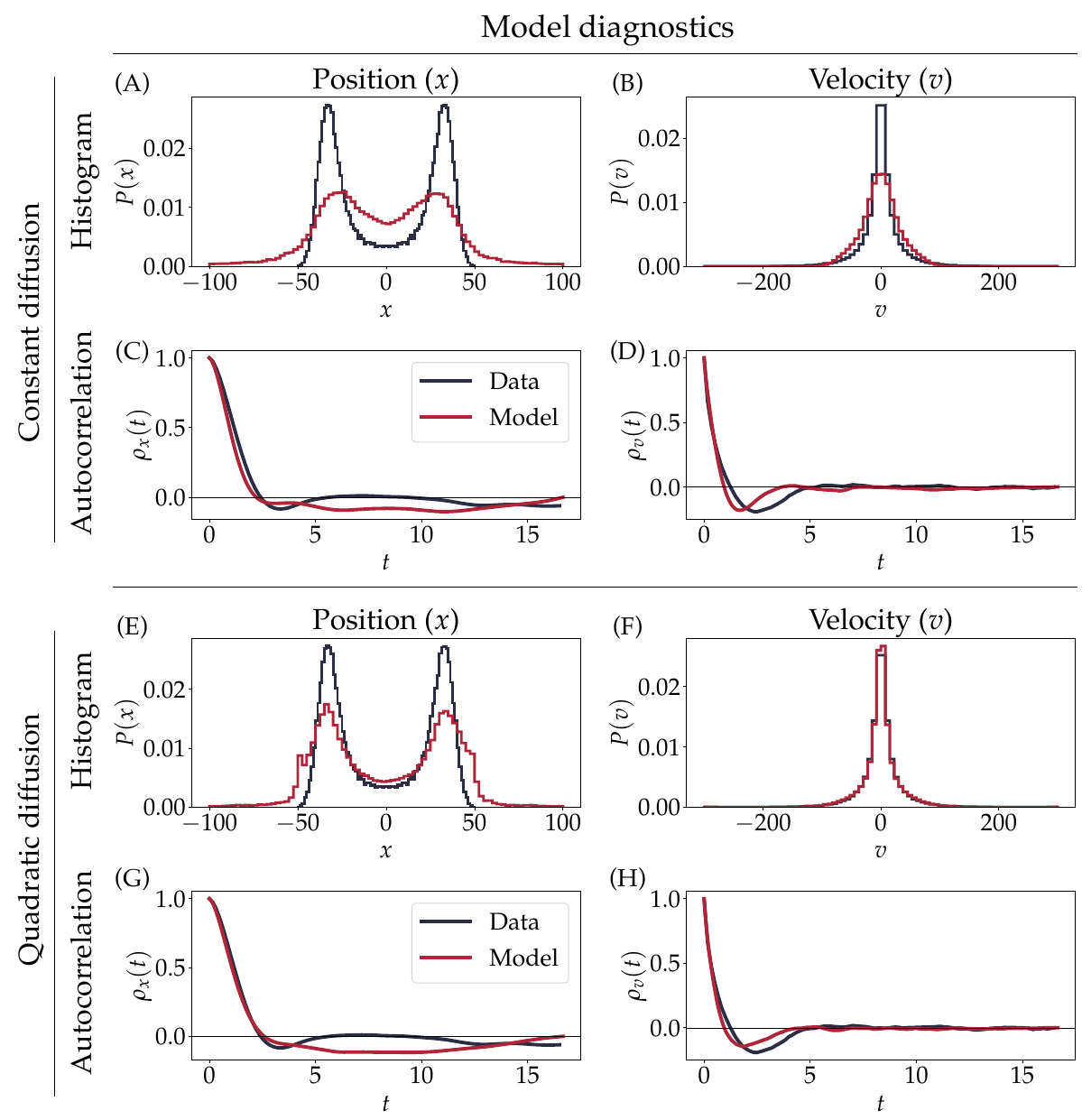} 
    \caption{Model diagnostics for an SDE model for cell migration, with constant, additive noise (A-D) and a quadratic multiplicative noise (E-H). Neither of these models is able to capture the distribution of $x$ correctly.
    }
    \label{fig:cell-mig-apx-diag}
\end{figure}

Since the overdamped model proved insufficient, we use an underdamped model as described in the main text. Below, we describe the detailed estimation procedure for an underdamped SDE for cell dynamics.

Recall that the STLSQ algorithm works by iteratively eliminating coefficients from a discovered function until we reach a minimal model. This procedure works correctly only when all the component variables ($x$ and $v$ in this case) are of the same scale. Therefore, as a preprocessing step, we rescale $x$ and $v$ to fall between $-1$ and $1$ based on the physical constraints of the system. Specifically, we define the non-dimensionalized variables $
\hat x = x / x_s$,  $\hat v = v / v_s$ and $\hat t = t / t_s$ where $x_s = 50 \, \mu\text{m}$, $v_s = x_s / \Delta t$ and $t_s = x_s / v_s$. ($\Delta t = 10$ min is the sampling interval.). We can now discover the following dimensionless SDE using the standard STLSQ algorithm:

\begin{align}
    \frac{d \hat v}{d \hat t} &= \hat f(\hat x, \hat v) + \hat g(\hat x, \hat v) \cdot \eta(t)
\end{align}

Based on visual examination, we conjecture that the drift function is of order 3 or higher, while the diffusion is small in magnitude but has a complex shape. We choose a cubic function for the drift and a quartic (4th order) polynomial for the diffusion. This combination is able to generate simulated time series that reasonably match the histograms and autocorrelation functions of $x$ and $v$ (Fig.5). Note that the the quartic function is essential to characterize the structure of the drift function. Neither a constant additive noise, nor a lower order polynomial function was able to produce a model consistent with the observed distribution of $x$ (Fig.~\ref{fig:cell-mig-apx-diag}).

Finally, we need to rescale the fitted SDE to the the original, dimensional variables $x$ and $v$. Let the non-dimensional drift and diffusion functions be $\hat f( \hat x, \hat v)$ be $\hat g(\hat x, \hat v)$ respectively. Then, the actual, dimensional $f$ and $g$ are given as:

\begin{align}
    f(x, v) &= \frac{v_s}{t_s} \cdot \hat f \left( \frac{x}{x_s}, \frac{v}{v_s} \right) \\
    g(x, v) &= \frac{v_s}{t_s} \cdot \hat g \left( \frac{x}{x_s}, \frac{v}{v_s} \right)
\end{align}

Tables \ref{tab:f_coeffs} and \ref{tab:g_coeffs} show the coefficients of the estimated drift and diffusion functions, for both the dimensionless and rescaled dimensional coordinates.

\newcommand{\EE}[1]{ \times 10^{#1}}

\begin{table}[]
    \centering
    \begin{tabular}{crr}
        \toprule
         \textbf{Term}   & \textbf{Coeff. in $\hat f$} & \textbf{Coeff. in $f$}  \\ 
         \midrule
         $x$    & -0.035    & $-1.272$\\ 
         $x^2$  & 0.038     & $5.510 \EE{-4}$\\ 
         $v$    & 0.083     & $49.84$\\ 
         $x^2v$ & -1.458    & $-3.499 \EE{-3}$\\ 
         $xv^2$ & -3.723    & $-1.489 \EE{-3}$\\ 
         $v^3$  & -3.019    & $-2.012 \EE{-4}$\\ 
         \bottomrule
    \end{tabular}
    \caption{The fitted coefficients for the drift function, in the dimensionless coordinates ($\hat f$) and the rescaled, dimensional coordinates ($f$).}
    \label{tab:f_coeffs}
\end{table}

\begin{table}[]
    \centering
    \begin{tabular}{crr}
        \toprule
         \textbf{Term}       & \textbf{Coeff. in $\hat g^2$} & \textbf{Coeff. in $g^2$}  \\
         \midrule
         $1$        &  0.010    & $ 3.195 \EE{4}$          \\
         $x^2$      & -0.023    & $-1.197 \EE{-2}$    \\
         $x^4$      &  0.015    & $ 1.259 \EE{-9}$           \\
         $xv$       & -0.232    & $-3.341 \EE{-3}$   \\
         $x^3v$     &  0.283    & $ 6.529 \EE{-10}$   \\
         $v^2$      & -0.129    & $-5.159 \EE{-5}$   \\
         $x^2v^2$   &  1.270    & $ 8.131 \EE{-11}$   \\
         $xv^3$     &  1.476    & $ 2.624 \EE{-12}$   \\
         $v^4$      &  0.582    & $-2.874 \EE{-14}$   \\
         \bottomrule
    \end{tabular}
    \caption{The fitted coefficients for the diffusion function, in the dimensionless coordinates ($\hat g^2$) and the rescaled, dimensional coordinates ($g^2$).}
    \label{tab:g_coeffs}
\end{table}

\subsection{Generalizability of the discovered SDE models}
\begin{figure*}
    \centering
    \includegraphics[width=\textwidth]{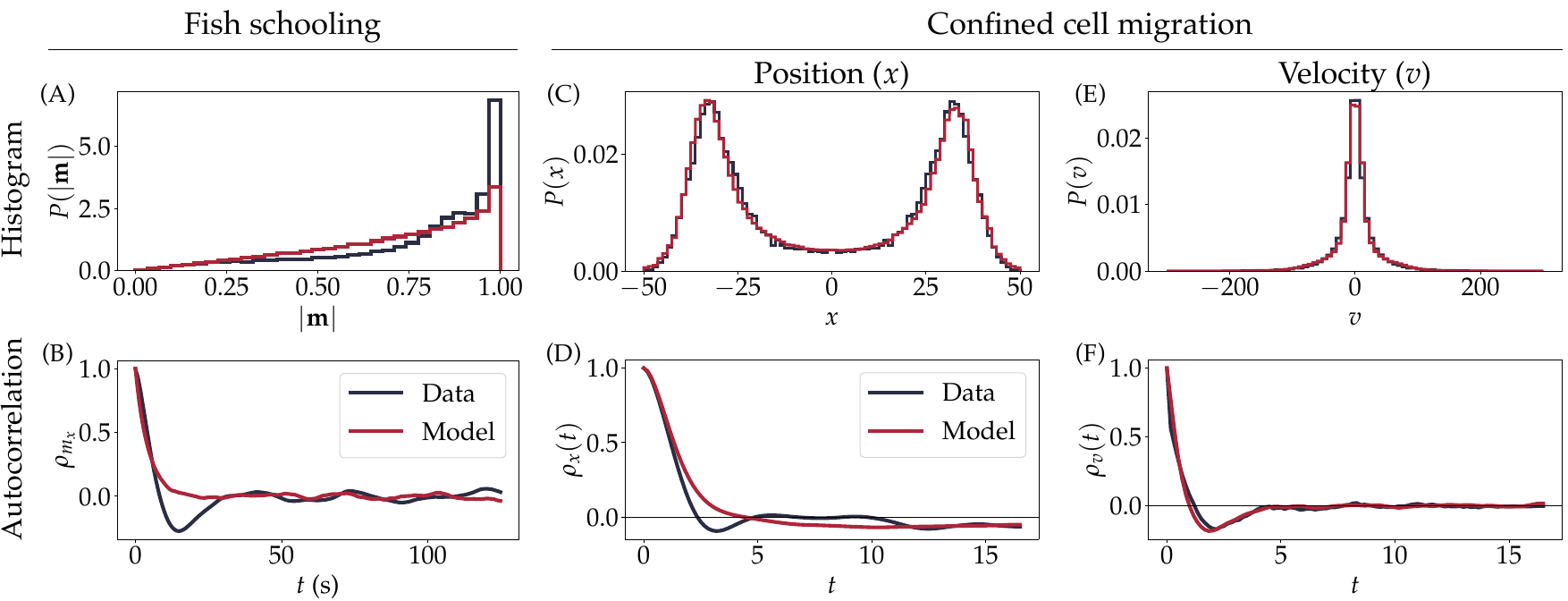} 
    \caption{\textbf{Testing generalizability of the discovered model with left-out data.} (A, B) Histogram and autocorrelation function of an SDE for the fish schooling dataset, estimated with only half of the data, compared to actual histogram and autocorrelation of the remaining half of the data. (C-F) Similarly, histogram and autocorrelation of $x$ and $v$ from the SDE model estimated with half of the data, compared the remaining half of the data.
    The estimated models show a very good match to the statistics of the left-out data.}
    \label{fig:apx-diag-leftout-data}
\end{figure*}

To confirm that the SDE discovery approach discovers generalizable models, we validate the discovered with left-out data (see subsection \emph{Diagnostics} -- Validation with left-out data) for both the fish and cell datasets. Here, we divide the datasets into two halves, a \emph{training set} and a \emph{validation set}. For the fish schooling dataset, two out of four available trials is used as the training set. For the cell migration dataset, 75 out of the 149 cell trajectories are used. We perform the model discovery procedure on the training set. We can now compute the compute the probability histogram and autocorrelation for a time series generated by the discovered model and compare these with the histogram and autocorrelation of the left-out data. For both the fish schooling and the cell migration dataset, the statistics of the model-generated time series match those that of the validation data (Fig~\ref{fig:apx-diag-leftout-data}), suggesting that the discovered SDEs have captured generalizable features of the underlying dynamics.

\section{Pitfalls and limitations}
\subsection{Estimation with limited data}
\label{ax:limited-data}

In this section, we examine the effects of data limitations on the estimation performance. Two aspects are considered, namely, the length of the time series, and the sampling interval.

\begin{figure*}
    \centering
    \includegraphics[width=\textwidth]{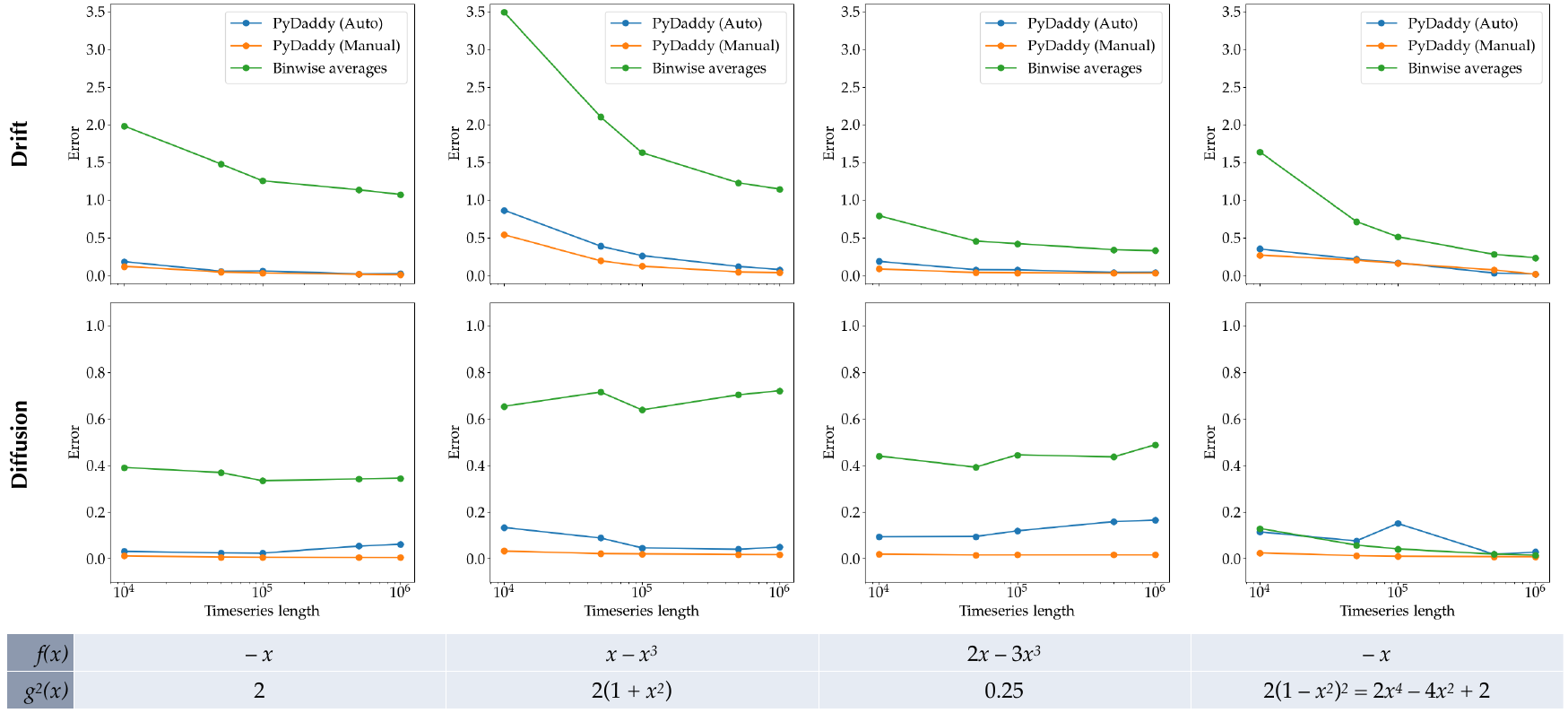} 
    \caption{\textbf{Effect of time series length on estimation error.} Estimation errors (relative r.m.s. error) for drift (top) and diffusion (bottom) for four example SDEs. The actual drift and diffusion functions are shown at the bottom. Estimation errors are plotted for conventional bin-wise estimation, PyDaddy estimation with \textit{auto}matic threshold tuning, and PyDaddy estimation with \textit{manual} thresholds.
    In general, as the amount of available data (i.e. time series length) increases, estimation error of drift decreases. Estimation accuracy for diffusion is good even with shorter time series.
    For both drift and diffusion, PyDaddy achieves much better estimation accuracy than conventional bin-wise averaging, even for short time series.
    }
    \label{fig:length}
\end{figure*}

\paragraph*{Effect of time series length}
In general, one would expect the estimation performance to increase when we have more data. For very short time series, the estimates are noisy and the estimation error is relatively high. As the amount of available data increases, the estimation error decreases.

Fig.~\ref{fig:length} shows the average estimation error, with simulated data of different lengths (ranging from $10^4$ to $10^6$ time points), for different models. The averages are computed over 100 different instantiations of each model. As expected, the estimation error for drift decreases with increasing time series length. For diffusion, the estimates are fairly accurate even with short time series, and there is no significant decrease in the estimation error with longer time series.

For comparison, the estimation errors of the bin-wise averaging method are also shown. In general, the estimation errors with the bin-wise averages are much larger than PyDaddy estimation errors. This is due to the fact that the bin-wise approach estimates drift and diffusion functions independently for each bin, and disregards the overall shape and smoothness of the functions. Errors in estimates for individual bins therefore can become quite large when there are only a few samples in the bin. By fitting an analytic function to the jump moments, PyDaddy is able to capture the smoothness of the drift and diffusion functions to achieve better estimates.

\begin{figure*}
    \centering
    \includegraphics[width=\textwidth]{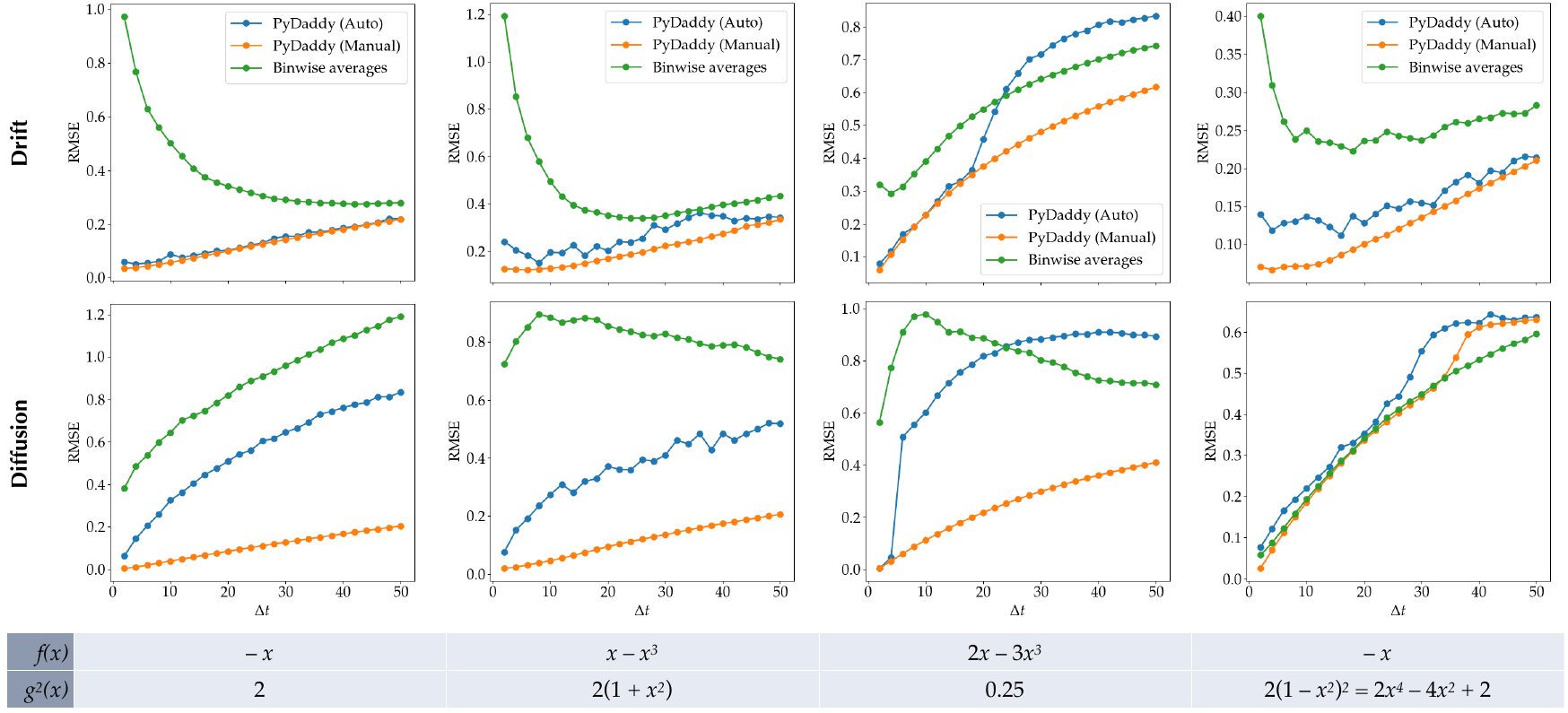} 
    \caption{\textbf{Effect of sampling time on estimation error.} Estimation errors (relative r.m.s. error) for drift (top) and diffusion (bottom), as a function of sampling interval $\Delta t$, for four example SDEs. The actual drift and diffusion functions are shown at the bottom. Estimation errors are plotted for conventional bin-wise estimation, PyDaddy estimation with automatic threshold tuning, and PyDaddy estimation with manual thresholds.
    For PyDaddy estimates, the estimation error always increases monotonically with subsampling time $\Delta t$. Contrast this with the bin-wise estimates, where the estimation error often has a non-monotonic trend.
    }
    \label{fig:deltat}
\end{figure*}

\paragraph*{Effect of sampling interval}
In the conventional approach of estimating drift and diffusion functions as bin-wise averages, there is a bias-variance trade-off: as the sampling interval increases, the bias in the estimation of drift and diffusion increases, but the variance decreases. To obtain the best trade-off, one may need to subsample the time series at a $\Delta t$ much larger than the resolution at which the data is available in~\cite{jhawar2020inferring}. Fig.~\ref{fig:deltat} shows how, for the bin-wise estimates, the estimation error decreases to a minimum (due to decreasing variance) before starting to increase again as $\Delta t$ increases. This is due to high variance in the bin-wise estimates for small values of $\Delta t$. The effect is particularly evident in the drift estimate. 

However, this effect is absent in the sparse regression approach, and smaller $\Delta t$ always yields smaller estimation errors.The smallest possible $\Delta t$ (i.e., the temporal resolution at which the data is available in) is usually the best choice for estimation with PyDaddy, further subsampling only decreases estimation performance.

\subsection{Identifying unsatisfactory models using diagnostic tests}
\label{apx:diag}

\begin{figure*}
    \centering
    \includegraphics[width=\textwidth]{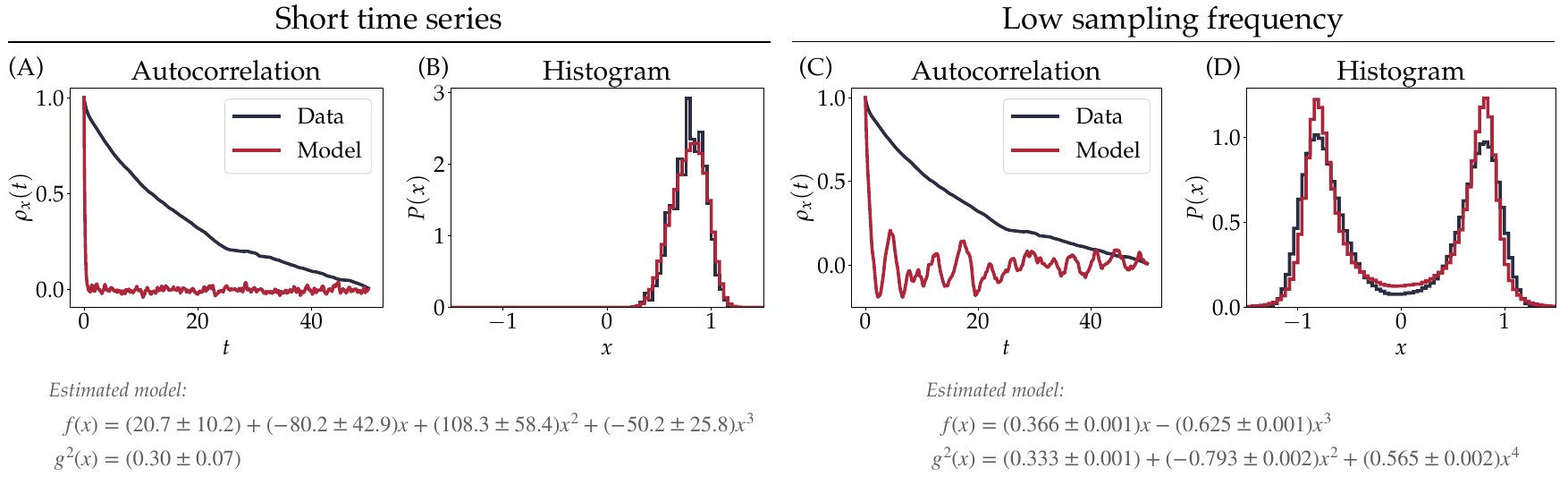} 
    \caption{\textbf{Estimating and diagnosing SDEs with insufficient data.} (A, B) Histogram and autocorrelation function of an SDE estimated with from a short time series (1000 time points). Both plots show a clear mismatch between the data and the model, suggesting that the estimated model is unreliable. (C, D) Histogram and autocorrelation function of an SDE estimated from a time series with a long sampling interval (1s). The autocorrelation function clear mismatch between the data and the model, suggesting that the estimated model is unreliable.}
    \label{fig:apx-diag-limited-data}
\end{figure*}

The section \emph{Pitfalls and how to avoid them} in the main text discusses situations in which the SDE discovery procedure can fail, and pointed out ways to spot these situations using diagnostic tests and---when possible---correct them. In this section, we explore each these situations in more detail, and demonstrate this using simulated datasets.

\begin{figure}
    \centering
    \includegraphics[width=0.6\linewidth]{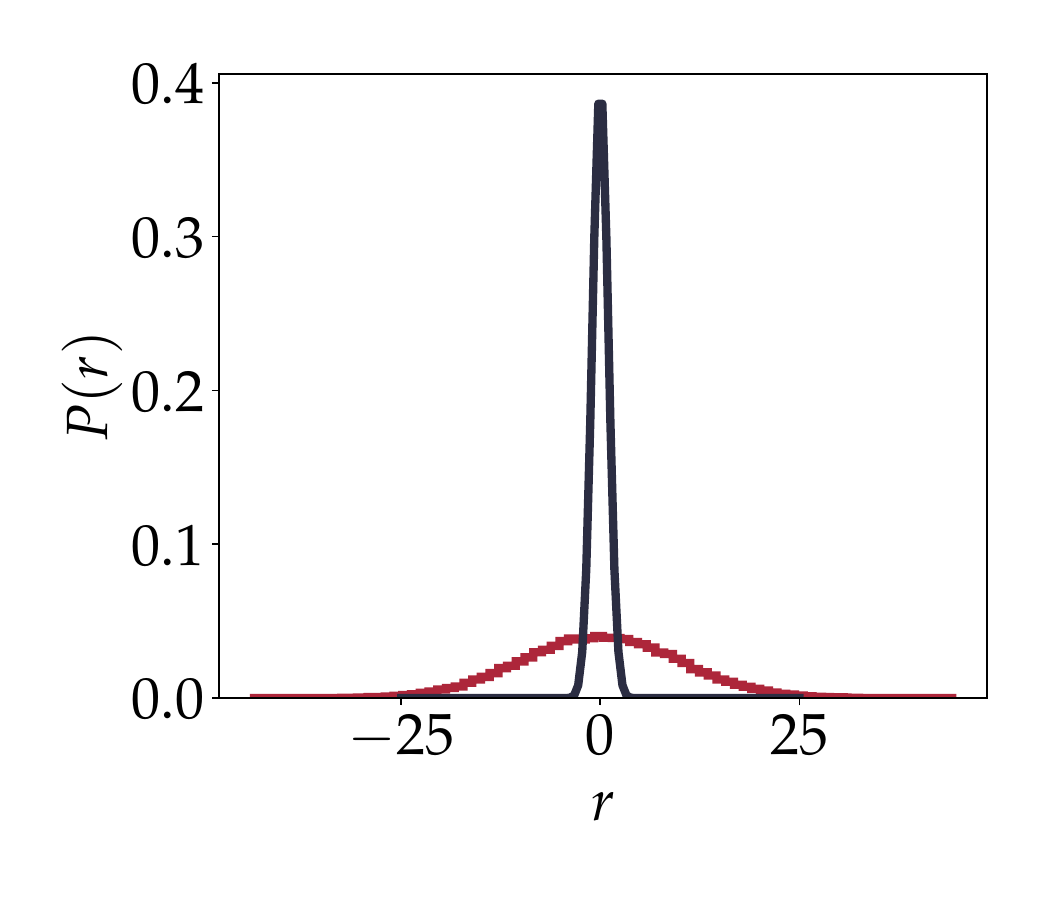} 
    \caption{Residual histogram for an SDE model estimated from the observation of a single species, where the actual dynamics follows a competition model with two interacting species. The estimated residual (red line) has a very large variance, and is not a standard normal distribution (black line) as expected. }
    \label{fig:apx-diag-lv-single}
\end{figure}

\paragraph*{Dimensionality of the system.} As described in the main text, there are often ways to reduce a high-dimensional system to a small number of coarse-grained variables, which are sufficient to describe the key dynamics we are interested in. The fish schooling dataset was a clear example of this, where we used a 2-dimensional polarization vector to describe the ordering dynamics of the school of fish, although the full dynamics of each individual fish can be far more complicated.

On the other hand, there can be situations where the actual dynamics of the system is high-dimensional, while we have access only to a few dynamical variables. In such situations, a model simply in terms of the observed variables will not be an appropriate model for the system dynamics. In these cases, we can use diagnostic tests to spot the wrongly identified model. As an example, consider the case of two interacting populations, as modelled by the Lotka-Volterra competition model (Eqn.~\ref{eq:lv-comp}), and suppose we have measured the population dynamics for one of the species. A single-variable SDE, capturing only the dynamics of the observed population, will be incomplete in this situation. Fig.~\ref{fig:apx-diag-lv-single} shows how the noise diagnostics can spot the misidentified model in this situation. As shown in the figure, the distribution of the residual is very different from the theoretical expectation of a standard normal distribution. This can be taken as an indicator that the identified SDE model is not the correct model and there are potential unexplained features in the residual.

The overdamped model (Eq.~\ref{eq:cell-eqn-overdamped} for the cell dynamics is another example. Again, we are attempting to model a system with a single variable, namely the cell position $x$, while a proper description requires the position and velocity. In this case, we saw that the residual time series had a finite autocorrelation time (Fig.~\ref{fig:cell-mig-overdamped}D), denoting that the underdamped model is not sufficient to capture the full dynamics of the system~\ref{fig:cell-mig-overdamped}.

\paragraph{State-space coverage} 
It is also important that the observed time series data is a good representation of the underlying dynamics of the system. For this, the time series should sufficiently explore the state-space of the system. For instance, if a system with multiple stable states is stuck near one of its stable points, we may not be able to discover a model that correctly captures the full phase portrait (see subsection, \emph{Inaccurate model discovery with limited data} for a demonstration). 

In many situations, this may not be a problem and a \emph{local} model of the dynamics may be good enough. In addition, in systems where experimental manipulations are possible, one can design different treatments of experiments to generate system time series with different initial conditions. Each of these treatment may explore one part of the state-space and converge to a locally stable state. By pooling together many such treatments and using the pooled data for estimation, one can obtain an SDE that describes the global dynamics of the system. This approach is demonstrated for some of the synthetic models in SI Section S3.

\paragraph{Limitations of sparse regression} 
Finally, the use of sparse regression for equation-learning can be perceived as somewhat limiting, as it limits the space of functions that can be discovered to those representable as linear combinations of the functions in the library (see Appendix A). However, the default choice of a polynomial basis as the library is suitable for many applications, as demonstrated with two contrasting datasets of cells and fish schools. Even for systems where non-polynomial approximations are required, a custom library can be specified in PyDaDDy—--indeed, this approach was used in some of the models in section \emph{Demonstration with cla with classical models in theoretical ecology} and SI Section S3. 

As an alternative, when more complicated non-linear functions are required, the jump moment discovery in PyDaDDy can be combined with other equation discovery techniques such as \emph{symbolic regression}~\cite{cranmer2023pysr,udrescu2020ai,tenachi2023deep}. Again, when the discovered model is not good enough---for instance, due to a poor choice of library---the model diagnostics can help spot problems and lead one to design a better candidate library.

\bibliography{references}